\documentclass[11pt]{book}

\usepackage{jfaa2e}     
\usepackage[latin1]{inputenc} 
\usepackage{amssymb}
\usepackage{amsfonts} 
\usepackage{graphicx} 
\usepackage{amsthm} 
\usepackage{amsmath,amscd} 
\usepackage{epsfig}

\def\ni{\noindent}
\def\nn{\nonumber}

\def \bc {\begin{center}}
\def \ec {\end{center}}
\def \bi {\begin{itemize}}
\def \ei {\end{itemize}}

\def \ba {\begin{array}}
\def \ea {\end{array}}

\def \bea {\begin{eqnarray}}
\def \eea {\end{eqnarray}}

\def \be {\begin{equation}}
\def \ee {\end{equation}}

\def \sd {\mathbb{S}^2}
\def \ca {{\cal A}}
\def \cf {{\cal F}}

\def \cc {{\cal C}}
\def \ct {{\cal T}}
\def \cb {{\cal B}}

\def \cn {{\cal N}}
\def \um {\frac{1}{2}}
\def \bz {\bar{z}}

\newtheorem{thm}{Theorem}[section]
\newtheorem{cor}[thm]{Corollary}
\newtheorem{lem}[thm]{Lemma}
\newtheorem{prop}[thm]{Proposition}

\theoremstyle{remark}
\newtheorem{rem}[thm]{Remark}

\newcommand{\li}{\langle}
\newcommand{\ld}{\rangle}

\begin{document}

\author{M. Calixto, J. Guerrero and J.C. Sánchez-Monreal}


\chapter{Sampling Theorem and Discrete Fourier Transform on the
Riemann Sphere}

\footnotetext{\textit{Math Subject Classifications.}
                    32A10, 42B05, 94A12, 94A20,  81R30.}

\footnotetext{\textit{Keywords and Phrases.}
                     holomorphic functions, Coherent States, Discrete Fourier Transform,
             Sampling, Frames}

\begin{abstract}
Using coherent-state techniques, we prove a sampling theorem for
Majorana's (holomorphic) functions on the Riemann sphere and we provide an
exact reconstruction formula as a convolution product of $N$ samples and a
given reconstruction kernel (a sinc-type function). We also discuss the
effect of over- and under-sampling. Sample points are roots of unity, a
fact which allows explicit inversion formulas for resolution and
overlapping kernel operators through the theory of Circulant Matrices and
Rectangular Fourier Matrices. The case of band-limited functions
on the Riemann sphere, with spins up to $J$, is also considered. The
connection with the standard Euler angle picture, in terms of spherical
harmonics, is established through a discrete Bargmann transform.
\end{abstract}

\section{Introduction}

The Fourier transform on the sphere is applied in a wide variety of fields:
geophysics, seismology, tomography, atmospheric science, computer vision,
atomic physics, astrophysics, statistics, signal processing,
crystallography, etc. It is therefore of great interest to develop
efficient techniques for the computation of Fourier coefficients,
spherical convolutions, etc..

Sometimes we have at our disposal just a set of samples of our signal and
we ask ourselves whether the Fourier transform may be computed, or the
whole signal be reconstructed (up to a certain degree of accuracy), from
the discrete samples. In the case of band-limited functions on the line
(or Abelian harmonic analysis in general), the classical (Shannon)
sampling theorem provides the necessary and sufficient conditions for this
problem. However, the establishment of sampling theorems for harmonic
analysis on non-Abelian groups and their homogeneous spaces is still
relatively scarce in the literature, apart from some important general
results for compact groups \cite{M1,M2} and the (noncompact) motion group
\cite{Ch3}. Moreover, we would want our algorithms to be \emph{fast} and
efficient. The Fast Fourier Transform (FFT), in the setting of Abelian
harmonic analysis (i.e., the well known Cooley-Tukey algorithm
\cite{Cooley} for time series analysis), has been extensively studied in
both the theoretical and applied literature but, again, there are few
algorithms for the efficient computation of Fourier transforms associated
with non-Abelian groups and their homogeneous spaces (see again Refs.
\cite{M1,M2} for compact groups and  \cite{Ch1} for the motion group and
its engineering applications \cite{Ch3}, namely in robotics \cite{Ch2}).
For finite non-Abelian groups, like the symmetric group $S_n$, the
reference \cite{DR} provides efficient algorithms to compute Fourier
transforms.

For the two-dimensional sphere $\mathbb{S}^2$, the efficient computation
of Fourier transforms of band-limited functions (those functions in
$L^2(\mathbb{S}^2)$ which expansion requires only spherical harmonics of
angular momentum at most $J$) has been achieved in, see for instance,
Refs. \cite{DH,MR,R,HRKM}. In reference \cite{DH}, the authors develop a
sampling theorem on the sphere, which reduces the computation of Fourier
transforms and convolutions of band-limited functions to discrete (finite)
calculations. Here, band-limited functions on $\sd$, of bandwidth $J$, are
expanded in terms of spherical harmonics and sampled at an equiangular
grid of $4J^2$ points.

The point of view followed in these references is a group theoretic one.
In this setting, the FFT on $\sd$ is an algorithm for the efficient
expansion of a function defined on the sphere $\sd=SO(3)/SO(2)$ in terms
of a set of irreducible matrix coefficients for the  special orthogonal
group in three dimensions, $G=SO(3)$, which,  in this case, are the
standard family of spherical harmonics.

In this article we consider the group $G=SU(2)$ (double cover of $SO(3)$),
which allows for (extra) half-integer angular momenta (spin). Moreover, we
shall work in a different (holomorphic) picture and use, instead of
spherical harmonics (based on an Euler angle characterization), another
system of (less standard) orthogonal polynomials: ``Majorana's
(holomorphic) functions'' \cite{Majorana,Dennis} on the Riemann sphere
$\bar{\mathbb{C}}=\mathbb{C}\cup\{\infty\}$ (one-point compactification of
the complex plane). The advantage of using this ``complex holomorphic
picture'', instead of the standard ``Euler angle picture'', is twofold:
firstly we can take advantage of the either diagonal or circulant
structure of resolution and overlapping kernel operators, respectively, to
provide explicit inversion formulas and, secondly, we can extend the
sampling procedure to half-integer angular momenta $s$, which could be
useful when studying, for example, discrete frames for coherent states of
spinning particles in Atomic Physics (see e.g. Refs. \cite{Klauder,CS} for
a thorough exposition on coherent states and its applications in Physics).
Moreover, for integer angular momenta $s=j$, we could always pass from one
picture to another through the Bargmann transform (\ref{Bargmann}).

Working with a fixed angular momentum (spin) $s$, we shall introduce a
system of coherent states for $SU(2)$ (the spin coherent states), which is
a set of states sharing similar properties with wavelets (in fact, they
can be considered the same thing, see \cite{Gazeau,Fuhr}). We shall
provide a generalized Bargmann Transform \cite{Bargmann} relating both
pictures (representations): the ``holomorphic'' one and the ``standard''
one, which is a particular case of coherent-state transform
\cite{CS,Klauder}. Then we shall choose in $\bar{\mathbb{C}}$ the roots of
unity as sampling points, so that the sampling of the coherent-state
overlap (or Reproducing Kernel) has a ``circulant'' structure
\cite{circulante}. Using the properties of the Rectangular Fourier
Matrices (RFM) and the theory of Circulant Matrices we will be able to
invert the (sampled) reproducing kernel $\cb$ and provide a reconstruction
formula for Majorana's (holomorphic) functions on the Riemann sphere. The
inversion formula is accomplished through an eigen-decomposition
$\cb={\cal F}D {\cal F}^{-1}$ of $\cb$, where ${\cal F}$ turns out to be
the standard discrete Fourier transform matrix. This fact allows for a
straightforward fast extension of the reconstruction algorithm. The case
of band-limited functions is also considered, but in this case the
inversion should be done numerically, and no fast algorithm is available,
for the moment.

In order to keep the article as self-contained as possible, we shall
introduce in the next two sections general definitions and results about
coherent states and frames based on a group $G$, and the standard
construction of spin coherent states for the case $G=SU(2)$. We refer the
reader to Refs. \cite{CS,Klauder,Holschneider,Gazeau} for more
information. In Section \ref {sampling} we provide sampling theorems and
reconstruction formulas for Majorana's functions on the Riemann sphere, and
discuss the effect of over- and under-sampling and the analogies with
the so called ``covariant interpolation''. We also discuss the case of
band-limited functions, where a negative result is proved in the case of
sampling at roots of unity. A reconstruction theorem is provided for another
set of sampling points, but the inversion should be done numerically. In
Section \ref{comments} we provide explicit expressions (discrete Bargmann
transforms) which connect our ``complex holomorphic picture'' and the
standard ``Euler angle picture'', and we discuss some obstructions that
arise. Appendices A and B are devoted to a brief review of rectangular
Fourier matrices and circulant matrices, respectively.

\section{A brief on Coherent States and Frames}
\label{CSandFrames}

Let us consider a \emph{unitary} representation $U$ of a Lie group $G$ on
a Hilbert space $({\cal H},\langle \cdot|\cdot\rangle)$. Consider also the
space
$L^2(G,dg)$ of square-integrable complex functions $\Psi$ on $G$, where
$dg=d(g'g),\,\forall g'\in G$, stands for the left-invariant
Haar measure, which defines the scalar product
\be
\left(\Psi|\Phi\right)=\int_G\bar{\Psi}(g)\Phi(g)dg. \ee A non-zero
function $\gamma\in {\cal H}$ is called \emph{admissible}  (or a
\emph{fiducial} vector) if $\Gamma(g)\equiv \langle
U(g)\gamma|\gamma\rangle\in L^2(G,dg)$, that is, if \be
c_\gamma=\int_G\bar{\Gamma}(g)\Gamma(g)dg=\int_G|\langle
U(g)\gamma|\gamma\rangle|^2dg<\infty. \label{norm}\ee

Let us assume that the representation $U$ is \emph{irreducible}, and that
there exists a function $\gamma$ admissible, then a system of coherent
states (CS) of ${\cal H}$  associated to (or indexed by) $G$ is defined
as the set of functions in the orbit of $\gamma$ under $G$
\be
\gamma_g=U(g)\gamma, \;\; g\in G.
\ee

 We can also restrict ourselves to a suitable
homogeneous space $Q=G/H$, for some closed subgroup $H$. Then, the
non-zero function $\gamma$ is said to be admissible mod$(H,\sigma)$ (with
$\sigma:Q\to G$ a Borel section), and the representation $U$ square
integrable mod$(H,\sigma)$, if the condition
\be
\int_Q|\langle U(\sigma(q))\gamma|\psi\rangle|^2 d q<\infty,\;\;\forall
\psi\in {\cal H}\label{qsquare}\ee holds, where $d q$ is a measure on $Q$
``projected'' from the left-invariant measure $dg$ on the whole $G$. The
coherent states indexed by $Q$ are defined as
$\gamma_{\sigma(q)}=U(\sigma(q))\gamma, q\in Q$, and they form an
overcomplete set in ${\cal H}$.

The condition (\ref{qsquare}) could also be written as an ``expectation
value" \be 0<\int_Q |\langle U(\sigma(q))\gamma|\psi\rangle|^2 dq=\langle
\psi|A_\sigma |\psi\rangle <\infty ,\;\;\forall \psi\in {\cal
H},\label{pbiop}\ee where
$A_\sigma=\int_Q|\gamma_{\sigma(q)}\rangle\langle \gamma_{\sigma(q)}|dq$
is a positive, bounded, invertible operator.\footnote{In this paper we
shall extensively use the Dirac notation in terms of ``bra'' and ``kets''
(see e.g. \cite{acha,Gazeau}). The Dirac notation is justified by the Riesz
Representation Theorem, and is valid in more general settings than Hilbert
spaces of square integrable functions
.}

If the operator $A_\sigma^{-1}$ is also bounded,
then the set $S_\sigma=\{|\gamma_{\sigma(q)}\rangle, q\in Q\}$ is called a
\emph{frame}, and a \emph{tight frame} if $A_\sigma$ is a positive
multiple of the identity, $A_\sigma=\lambda {I}, \lambda>0$.

To avoid domain problems in the following, let us assume that $\gamma$
generates a frame (i.e., that $A_\sigma^{-1}$ is bounded). The \emph{CS
map} is defined as the linear map \be\begin{array}{cccc} T_\gamma: & {\cal
H}&\longrightarrow&
L^2(Q,dq)\\
 & \psi & \longmapsto & \Psi_\gamma(q)=[T_\gamma\psi](q)=\frac{\langle
\gamma_{\sigma(q)}|\psi\rangle}{\sqrt{c_\gamma}},\end{array}
\label{cwt}.\ee  Its range $L^2_\gamma(Q,dq)\equiv T_\gamma({\cal H})$ is
complete with respect to the scalar product
$(\Phi|\Psi)_\gamma\equiv\left(\Phi|T_\gamma A_\sigma^{-1}
T_\gamma^{-1}\Psi\right)_Q$ and $T_\gamma$ is unitary from ${\cal H}$ onto
$L^2_\gamma(Q,dq)$. Thus, the inverse map $T_\gamma^{-1}$ yields the
\emph{reconstruction formula}
\be
\psi=T_\gamma^{-1}\Psi_\gamma=\int_Q\Psi_\gamma(q)A_\sigma^{-1}\gamma_{\sigma(q)}
d q,\;\;\Psi_\gamma\in L^2_\gamma(Q,d q),\ee which expands the signal
$\psi$ in terms of CS $A_\sigma^{-1}\gamma_{\sigma(q)}$ with wavelet
coefficients $\Psi_\gamma(q)=[T_\gamma\psi](q)$. These formulas acquire a
simpler form when $A_\sigma$ is a multiple of the identity, as is for the
case considered in this article.

When it comes to numerical calculations, the integral
$A_\sigma=\int_Q|\gamma_{\sigma(q)}\rangle\langle \gamma_{\sigma(q)}|d q$
has to be discretized, which means to restrict ourself to a discrete
subset ${\cal Q}\subset Q$. The question is whether this restriction will
imply a loss of information, that is, whether the set ${\cal
S}=\{|q_k\rangle\equiv|\gamma_{\sigma(q_k)}\rangle, q_k\in{\cal Q} \}$
constitutes a discrete frame itself, with resolution operator
\be {\cal
A}=\sum_{q_k\in {\cal Q}}|q_k\rangle\langle q_k|.\ee The operator ${\cal
A}$ need not coincide with the original ${A}_\sigma$. In fact, a
continuous tight frame might contain discrete non-tight frames, as happens
in our case (see later on Sec. \ref{sampling}).

Let us assume that ${\cal S}$ generates a discrete frame, that is, there
are two positive constants $0<b<B<\infty$ (\emph{frame bounds}) such that
the admissibility condition \be b||\psi||^2\leq|\sum_{q_k\in {\cal Q}}
\langle q_k|\psi\rangle|^2\leq B||\psi||^2\label{pbiop2}\ee holds $\forall
\psi\in{\cal H}$. To discuss the properties of a frame, it is convenient
to define the frame (or sampling) operator ${\cal T}:{\cal H}\to \ell^2$
given by ${\cal T}(\psi)=\{\langle q_k|\psi\rangle, \,q_k\in{\cal Q}\}$.
Then we can write ${\cal A}={\cal T}^*{\cal T}$, and the admissibility
condition (\ref{pbiop2}) now adopts the form \be b I\leq {\cal
T}^*{\cal T}\leq B I,\ee where $I$ denotes the identity operator in
${\cal H}$. This implies that ${\cal A}$ is invertible. If we define the
\emph{dual frame} $\{|\tilde{q}\rangle\equiv {\cal A}^{-1} |q\rangle\}$,
one can easily prove that the expansion (\emph{reconstruction formula})
\be |\psi\rangle=\sum_{q_k\in {\cal Q}}\langle
q_k|\psi\rangle|\tilde{q}_k\rangle\ee converges strongly in ${\cal H}$,
that is, the expression \be{\cal T}_l^+{\cal T}= \sum_{q_k\in {\cal
Q}}|\tilde{q}_k \rangle\langle q_k|= {{\cal T}}^*({\cal T}_l^+)^*=
\sum_{{q}_k\in {\cal Q}}|q_k \rangle\langle \tilde{q}_k |=
I\label{resolucionidentidad}\ee
provides a resolution of the identity, where
${\cal T}_l^+\equiv({\cal T}^*{\cal T})^{-1}{\cal T}^*$ is the (left)
pseudoinverse (see, for instance, \cite{pseudoinverse}) of ${\cal T}$ (see e.g. \cite{Holschneider,Gazeau}
for a proof, where they introduce the dual frame operator $\tilde{\ct}=(\ct_l^+)^*$ instead).

It is interesting to note that the operator $P={\cal T}{\cal T}_l^+$ acting
on $\ell^2$ is an orthogonal projector onto the range
of $\ct$.


We shall also be interested in cases where there are not enough points to
completely reconstruct the signal, i.e., \emph{undersampling}, but a
partial reconstruction is still possible. In these cases ${\cal S}$ does not generate a discrete frame,
and the resolution operator ${\cal A}$ would not be invertible. But we can
construct another operator from ${\cal T}$, ${\cal B}={\cal T}{\cal T}^*$,
acting on $\ell^2$.

The matrix elements of ${\cal B}$ are ${\cal B}_{kl}=\langle q_k|q_l\rangle$, therefore
${\cal B}$ is the discrete reproducing kernel operator, see eq. (\ref{CSoverlap}).
If the set ${\cal S}$ is linearly independent, the
operator ${\cal B}$ will be invertible and a (right)
pseudoinverse can be constructed for ${\cal T}$,
${\cal T}_r^+\equiv {\cal T}^*({\cal T}{\cal T}^*)^{-1}$, in such a way that
$\ct \ct_r^+ = I_{\ell^2}$. As in the previous case there is another operator,
$P_{\cal S}= \ct_r^+ \ct$ acting on ${\cal H}$ which is an orthogonal projector
onto the subspace spanned by ${\cal S}$. A pseudo-dual frame can be defined as
\be
|\tilde{q}_k\rangle = \sum_{q_l\in {\cal Q}} ({\cal
B}^{-1})_{lk}|q_l\rangle \ee
providing a resolution of the projector $P_{\cal S}$,
\be
 {\cal T}_r^+ \ct = \sum_{q_k\in {\cal Q}}|\tilde{q}_k \rangle\langle q_k|=
\ct^* ({\cal T}_r^+)^*  = \sum_{{q}_k\in {\cal Q}}|q_k \rangle\langle \tilde{q}_k |=
P_{\cal S} \label{resolucionproyector}
\ee
Using this,
an ``alias" $|\hat{\psi}\rangle$ of the signal $|\psi\rangle$ is obtained,
\be |\hat{\psi}\rangle=\sum_{q_k\in
{\cal Q}}\langle q_k|\psi\rangle|\tilde{q}_k\rangle
\ee
which is the orthogonal projection of $|\psi\rangle$ onto the subspace spanned by ${\cal S}$, $|\hat{\psi}\rangle=P_{\cal S}|\psi\rangle$. An example of this can be found in
Sec. \ref{undersampling}.

The two operators ${\cal A}$ and ${\cal B}$ are intertwined by the frame
operator ${\cal T}$, ${\cal T}{\cal A}={\cal B}{\cal T}$. If ${\cal T}$ is
invertible, then both ${\cal A}$ and ${\cal B}$ are invertible and ${\cal
T}_r^+={\cal T}_l^+ ={\cal T}^{-1}$. This case corresponds to critical
sampling, where both operators ${\cal A}$ and ${\cal B}$ can be used to
fully reconstruct the signal.

It should be noted that in the case in which there is a finite number $N$ of
sampling points $q_k$, the space $\ell^2$ should be substituted by
$\mathbb{C}^N$, and the operator $\cb$ can be identified with its matrix
once a basis has been chosen. If the Hilbert space ${\cal H}$ is finite
dimensional, as it is the case for all irreducible and unitary
representations of $SU(2)$, all operators appearing in this section
can be identified with their matrices.

\section{Representations of $SU(2)$: Spin Coherent States}

The subject of Harmonic Analysis on the rotation group has been
extensively treated in the literature. Here we shall try to summarize what
is important for our purposes, in order to keep the article as
self-contained as possible.

\subsection{Coordinate Systems and Generators}

The (two-dimensional) fundamental representation of the Lie group $SU(2)$
corresponds to the group of complex
$2\times 2$ unitary matrices with determinant one:
\be
SU(2)=\{ U(\zeta)=\left(
\begin{array}{cc}\zeta_1&\zeta_2\\-\bar{\zeta}_2&\bar{\zeta}_1\end{array}\right), \,\,
\zeta_{1},\zeta_{2} \in {\mathbb C}:
\det(U)=|\zeta_{1}|^{2}+|\zeta_{2}|^{2}=1 \}. \ee
The coordinates $\zeta_1, \zeta_2$ are called ``Cayley-Klein'' parameters
in the literature. Writing
\be
\zeta_{1}=\epsilon_{0}+i\epsilon_{3}, \,\,\,
\zeta_{2}=\epsilon_{2}+i\epsilon_{1}, \, \epsilon_j\in\mathbb R,\ee
we have that
\be
\det(U)=|\zeta_{1}|^{2}+|\zeta_{2}|^{2}=\epsilon_{0}^{2}+\epsilon_{1}^{2}+\epsilon_{2}^{2}+
\epsilon_{3}^{2}=1, \ee which tells us that $SU(2)\approx \mathbb S^{3}$
(the four-dimensional sphere) as a (three-dimensional) manifold. Denoting
by
\be
J_1=\frac{1}{2}
  \begin{pmatrix}
    0 & 1 \\
    1 & 0
  \end{pmatrix},\,J_2=\frac{1}{2}
  \begin{pmatrix}
    0 & -i \\
    i & 0
  \end{pmatrix},\,J_3=\frac{1}{2}
  \begin{pmatrix}
    1 & 0 \\
    0 & -1
  \end{pmatrix}, \label{paulimat}
\ee
a basis of $2\times 2$ traceless Hermitian (halved Pauli) matrices, we can
also write any matrix $U\in SU(2)$, in a compact way, as
\be
U(\epsilon)=\epsilon_{0}I+2i\sum_{k=1}^3\epsilon_k J_k,\label{Klein}\ee
where $I$ stands for the $2\times 2$ identity matrix. The matrices
(\ref{paulimat}) are also called the generators of infinitesimal (small)
transformations $U=I+i\varepsilon A,\,\varepsilon<<1$, since $UU^*=I$ and
$\det(U)=1$ imply (up to quantities of order two) that $A$ is a traceless
Hermitian matrix, that is, it can be written as $A=\sum_{k=1}^3a_k J_k$.
The Lie algebra of infinitesimal generators of $SU(2)$ is defined as the
(real) vector space $su(2)={\rm Span}\{J_1,J_2,J_3\}$ of traceless
Hermitian matrices satisfying the standard (angular momentum) commutation
relations (easy to check):
\be [J_1,J_2]=iJ_3,\,[J_2,J_3]=iJ_1, \,
[J_3,J_1]=iJ_2.\label{commutj}\ee
Any connected Lie group like $SU(2)$ can be built up by means of its
infinitesimal generators via the exponential:
\be
U(\alpha )=e^{i\sum_{k=1}^3\alpha_{k}
J_k}=\cos{\frac{\alpha}{2}}I+2i\sum_{k=1}^3n_k
\sin{\frac{\alpha}{2}}J_k\label{canonicalcoord} \ee where
$\alpha_k\in\mathbb R, k=1,2,3$, are called canonical coordinates at the
identity element $U=I$ and
$\alpha=\sqrt{\alpha_1^2+\alpha_2^2+\alpha_3^2}, \,\,
n_k=\frac{\alpha_k}{\alpha}$. Comparing (\ref{Klein}) with
(\ref{canonicalcoord}) gives a relation between the Cayley-Klein
parameters
$\epsilon$ and the canonical coordinates $\alpha$.

Let us introduce another complex parametrization of $SU(2)$, adapted to
the \emph{Hopf fibration} of $\mathbb S^3$, which will be of use in what
follows. Let us define the following equivalence relation in $SU(2)$:
\be
(\zeta'_{1},\zeta'_{2})\sim (\zeta_{1},\zeta_{2}) \Leftrightarrow
(\zeta'_{1},\zeta'_{2})=\eta (\zeta_{1},\zeta_{2});\,\,\eta\in\mathbb C,
|\eta|=1, \ee so that the quotient space (coset) $(SU(2)/{\sim})$
coincides with the complex projective space $\mathbb CP^{1}$, which is
isomorphic to
$\mathbb S^2$. Indeed, let us denote
by $[\zeta_{1},\zeta_{2}]$ an element (equivalence class) of $\mathbb
CP^{1}$. If
$\zeta_2\not=0$ then
$[\zeta_{1},\zeta_{2}]=[\frac{\zeta_{1}}{\zeta_{2}},1]=[z,1]$ represents a point $z\in \mathbb
C$, which is related to the stereographic projection of the Riemann sphere
on $\mathbb C$ (see later on this section). If
$\zeta_2=0$, then
$[\zeta_{1},0]=[1,0]$ is just a point (the north/south pole). The other
chart corresponds to $\zeta_1\not=0$, which contains the identity element
$U=I$ of $SU(2)$. We shall work in this chart and define
$z\equiv\frac{\zeta_{2}}{\zeta_{1}}$. The projection
\be \pi:SU(2)\rightarrow \mathbb S^{2}, \;(z_{1},z_{2})\mapsto
[z_{1},z_{2}]\ee
gives $SU(2)$ a principal fibre bundle structure with structural group

\be
\pi^{-1}([z_{1},z_{2}])=\{\eta\in\mathbb C: |\eta|=1 \}\simeq
U(1).\label{structuresub}\ee In our chart, we can take
$\eta=e^{i\varphi}=\frac{\zeta_{1}}{|\zeta_{1}|}$. The Cayley-Klein
parameters can be written in these Hopf-fibration coordinates as
\be
\zeta_{1}={\cal N}(z,\bar{z})\eta , \,\,\, \zeta_{2}={\cal
N}(z,\bar{z})z\eta;\;\; {\cal
N}(z,\bar{z})\equiv\sqrt{\frac{1}{1+z\bar{z}}},\label{cn}\ee
where we have defined the suitable normalization factor
${\cal N}$ for convenience. Denoting $J_±=J_1\pm iJ_2$ raising and
lowering ladder operators, we can check that any group element
$U\in SU(2)$ can also be written in complex coordinates $z,\eta$ as
\be
U(z,\bar{z},\varphi)=\cn(z,\bar{z}) e^{z J_-}e^{-\bz J_+}e^{-i\varphi
J_3}.\label{Hopfcomplex} \ee

We have discussed the (two-dimensional) fundamental representation of
$SU(2)$. There is also a three-dimensional (adjoint) representation of $SU(2)$
on its Lie algebra
\bea su(2)&=&\left\{ X=\sum_{k=1}^3 x_kJ_k=\frac{1}{2}
  \begin{pmatrix}
    x_3 & x_1-ix_2 \\
    x_1+ix_2 & -x_3
  \end{pmatrix}, x_k\in \mathbb R
\right\} \nn \\ &\simeq & \mathbb R^3=\left\{ (x_1,x_2,x_3), x_k\in\mathbb
R \right\}\eea given by the action
\be
U: su(2)\longrightarrow su(2),\,\, X\mapsto UXU^*,\label{adjointaction}\ee
which reduces to the standard action of the rotation group $SO(3)$, of
$3\times 3$ orthogonal matrices, on
$\mathbb R^3$. The fact that $U$ and $-U$ give the same rotation in  (\ref{adjointaction}) is a
consequence of the fact that $SO(3)=SU(2)/\mathbb Z_2$ or, in other words,
$SU(2)$ is the double cover of $SO(3)$. It is usual to parametrize $SO(3)$
in terms of Euler angles, which correspond to the choice (in the
arrangement
$x_3(\varphi)\to x_2(\theta)\to x_3(\phi)$)
\be
 U(\theta,\phi,\varphi)=e^{-i\phi J_3}e^{-i\theta J_2}e^{-i\varphi J_3}\label{Eulerangles}.
\ee
After a little bit of algebra (power expansion of the exponentials) , we
can find a relation between Cayley-Klein parameters and Euler angles given
by
\be \zeta_1=e^{i\frac{\varphi+\phi}{2}}\cos{\frac{\theta}{2}},\;
\zeta_2=e^{i\frac{\varphi-\phi}{2}}\sin{\frac{\theta}{2}},\ee
so that $z=\frac{\zeta_{2}}{\zeta_{1}}=e^{i\phi}\tan(\frac{\theta}{2})$ is
the stereographic projection of the Riemann sphere on the complex plane,
as anticipated before.

We have discussed the two-dimensional (spin $s=1/2$) and three-dimensional
(spin $s=1$) representations of $SU(2)$ in order to introduce coordinate
systems. Let us consider now higher-dimensional unitary irreducible
representations of arbitrary spin $s$.

\subsection{Higher-Spin Representations}

Unitary irreducible representations of the Lie algebra $su(2)$ are
$(2s+1)$-dimensional, where $s=0,1/2,1,3/2,\dots$ is a half-integer
parameter (spin or angular momentum) that labels each representation. Each
carrier space ${\cal H}_s\simeq \mathbb C^{2s+1}$ is spanned by the common
angular momentum orthonormal basis $B({\cal H}_s)=\{|s,m\rangle,
m=-s,\dots,s\}$ (in bra-ket notation) of eigenvectors of $J_3$ and the
Casimir (central) operator $\vec{J}^{\,2}=J_1^2+J_2^2+J_3^2$, i.e.,
\be
J_3|s,m\ld=m|s,m\ld,\,\,\vec{J}^{\,2}|s,m\ld=s(s+1)|s,m\ld.\label{repreang0}\ee
{}From the commutation relations
\be
[J_3,J_\pm]=\pm J_\pm\ee we see that $J_\pm$ play the role of raising and
lowering ladder operators, respectively, whose action on the basis vectors
proves to be
\be J_\pm|s,m\ld=\sqrt{(s\mp m)(s\pm m+1)}|s,m\pm 1\ld.\label{repreang}\ee
Indeed, it can be easily check that the action (\ref{repreang}) preserves
the commutation relations (\ref{commutj}); for example:
\be [J_+,J_-]|s,m\ld=\dots=2m|s,m\ld=2J_3|s,m\ld,\ee
and so on.

Note that the structure subgroup $U(1)\subset SU(2)$ in
(\ref{structuresub}), generated by
$J_3$, stabilizes any basis vector up to an overall multiplicative phase
factor (a character of $U(1)$), i.e., $e^{-i\varphi
J_3}|s,m\ld=e^{-im\varphi}|s,m\ld$. Thus, according to the general
prescription explained in Sec. \ref{CSandFrames}, letting
$Q=SU(2)/U(1)=\sd$ and taking the Borel section $\sigma:Q\to G$ with
$\sigma(\phi,\theta)=(\theta,\phi,\varphi=0)$, or
$\sigma(z,\bz)=(z,\bz,0)$, we shall define, from now on, families of
covariant coherent states ${\rm mod}(U(1),\sigma)$ (see \cite{Gazeau}). In
simple words, we shall set
$\varphi=0$ and drop it from the vectors: $U(\theta,\phi,\varphi)|s,m\ld$
and $U(z,\bar{z},\varphi)|s,m\ld$.

Therefore, we have different characterizations of spin coherent states
according to distinct choices of parameterizations. We shall concentrate
on the (Hopf) complex (\ref{Hopfcomplex}) and Euler angle
(\ref{Eulerangles}) parameterizations.

\subsection{Euler Angle Characterization: Spherical
Harmonics}\label{eacsh}

For any choice of fiducial vector $|\gamma\ld=|s,m\ld$ the set of coherent
states $|\theta,\phi;m\ld\equiv U(\theta,\phi)|\gamma\ld$ is overcomplete
(for any $m$) in ${\cal H}_s$. They can be easily computed by
exponentiating the relations (\ref{repreang0},\ref{repreang}). This set of
coherent states is also a tight frame with \be
A_\sigma=\frac{2s+1}{4\pi}\int_{\sd}|\theta,\phi;m\ld
\li\theta,\phi;m|d\Omega\ee a resolution of unity and $d\Omega=\sin\theta
d\theta d\phi$ the standard invariant measure on the 2-sphere. Indeed, due
to the invariance of the measure, it follows that $UA_\sigma=A_\sigma U$
for all
$U\in SU(2)$. Since the representation is irreducible we conclude from
Schur's Lemma that $A_\sigma=\lambda I$ for some constant $\lambda$.
Moreover,
${\rm Tr}(A_\sigma)=2s+1={\rm Tr}(I)\Rightarrow \lambda=1\Rightarrow A_\sigma=I$.

For the particular case of integer spin $s=j$ and fiducial vector $m=0$,
the standard \emph{spherical harmonics} $Y^m_j(\theta,\phi)$ arise as the
irreducible matrix coefficients (or Wigner ${\cal D}$-functions, see e.g.
Wigner's text book \cite{Wigner}):
\be \li \theta,\phi;0|j,m\ld=\li
j,0|U(\theta,\phi)^*|j,m\ld=\sqrt{\frac{4\pi}{2j+1}}Y^m_j(\theta,\phi),\label{spherical}\ee
or, in other words, the components of spin coherent states
$|\theta,\phi;0\ld$ over the orthonormal basis $\{|j,m\ld\}$. Thus, for a
general angular momentum $j$ state $|\psi\ld$ we have the standard
spherical harmonic decomposition [the wavelet coefficients (\ref{cwt})]
\be \Psi(\theta,\phi)=\li \theta,\phi;0|\psi\ld=
\sqrt{\frac{4\pi}{2j+1}}\sum_{m=-j}^{j}\psi_mY^m_j(\theta,\phi),\label{Wigner}\ee
with Fourier coefficients $\psi_m=\li j,m|\psi\ld$.

Spherical harmonics are rather well known special functions in the
literature. In this article, we shall work with a less standard set of
basis functions for the (complex) Riemann sphere: the Majorana functions.

\subsection{Complex Holomorphic Characterization: Majorana functions}

In this case we shall use $|\gamma\ld=|s,s\ld$ as fiducial vector (i.e.,
the highest weight vector), so that $J_+|\gamma\ld=0$ and the coherent
states
\be |z\ld\equiv U(z,\bz)|\gamma\ld=\cn_s(z,\bar{z})
e^{zJ_-}e^{-\bar{z}J_+}|s,s\ld=\cn_s(z,\bar{z}) e^{zJ_-}|s,s\ld,\ee
are holomorphic (only a function of
$z$), apart from the normalization factor
$\cn_s$ which, for higher-spin representations $s>1/2$, (slightly) differs from
$\cn$ in (\ref{Hopfcomplex}). In order to determine $\cn_s$, we first recall the relation
(\ref{repreang}) which, by exponentiation, gives
\be e^{zJ_-}|s,s\ld= |s,s\ld
+z\sqrt{2s}|s,s-1\ld+\frac{1}{2}z^2\sqrt{2s}\sqrt{2(2s-1)} |s,s-2\ld
+\dots+z^{2s} |s,-s\ld\equiv\cn_s^{-1}|z\ld.\label{exp-ansion}\ee Then,
imposing unitarity, i.e., $\li z|z\ld=1$, we arrive at $\cn_s=\cn^{2s}$,
with $\cn$ given in (\ref{cn}).

As for the Euler angle case, the frame $\{|z\ld, z\in\mathbb C\}$ is also
tight in
${\cal H}_s$, with resolution of unity
\be I=\frac{2s+1}{\pi}\int_{\mathbb C}|z\ld \li
z|\frac{d^2z}{(1+z\bz)^2},\label{resolholo}\ee where we denote
$d^2z=d{\rm Re}(z)d{\rm Im}(z)$. Indeed, using (\ref{exp-ansion}) we have that
\bea \frac{2s+1}{\pi}\int_{\mathbb C}|z\ld \li
z|\frac{d^2z}{(1+z\bz)^2}&=& \frac{2s+1}{\pi}\int_{\mathbb
C}\sum_{n,m=0}^{2s}\frac{z^n\bz^m}{n!m!}J_-^n|s,s\ld\li s,s|J_+^m\frac{
d{\rm Re}(z)d{\rm Im}(z)}{(1+z\bz)^{2s+2}}\\ &=& (2s+1)\sum_{n=0}^{2s}
\binom{2s}{n}\int_0^\infty |s,s-n\ld\li s,s-n|\frac{x^n
dx}{(1+x)^{2s+2}}\\ &=& \sum_{m=-s}^s|s,m\ld\li s,m|=I, \eea
where polar coordinates were used at intermediate stage. Also, the same
argument as in Sec. \ref{eacsh}, based on Schur's Lemma, is valid here.

Using (\ref{exp-ansion}), the decomposition of the coherent state $|z\ld$
over the orthonormal basis $\{|s,m\ld\}$ gives the irreducible matrix
coefficients
\bea \li z|s,m\ld &=&\li
s,s|U(z,\bz)^*|s,m\ld=\tbinom{2s}{s+m}^{1/2}(1+z\bz)^{-s}\bz^{s+m}\nn
\\ &\equiv& \cn(z,\bz)^{2s}\Upsilon^m_s(\bz),\label{upsilon}\eea where now
$\Upsilon^m_s(\bz)$ is just a monomial in $\bz$ times a numeric (binomial)
factor. A general spin $s$ state $|\psi\ld$ is represented in the present
complex characterization by the so called {\it Majorana function}
\cite{Majorana,Dennis}:
\be {\Psi}(z)\equiv\li
z|\psi\ld=(1+z\bz)^{-s}\sum_{m=-s}^{s}\psi_m\Upsilon^m_s(\bz)=\cn(z,\bz)^{2s}
f(\bz),\ee which is an anti-holomorphic function of $z$ (in this case, a
polynomial).\footnote{Here we abuse notation when representing the
non-analytic function $\Psi(z,\bar{z})$ simply as $\Psi(z)$, which is
indeed anti-holomorphic up to the normalizing, non-analytic (real),
pre-factor $\cn^{2s}=(1+z\bz)^{-s}$. Usually, this pre-factor is absorbed
in the integration measure in (\ref{resolholo}). If we choose the lowest
weight fiducial vector $|\gamma\ld=|s,-s\ld$, we would obtain proper
holomorphic functions $f(z)$.}

Note that the set of CS $\{|z\ld\}$ is not orthogonal. The CS overlap (or
Reproducing Kernel) turns out to be
\be C(z,z')=\langle
z|z'\rangle=\frac{(1+z'\bar{z})^{2s}}{(1+z\bar{z})^s(1+z'\bar{z}')^s}.\label{CSoverlap}
\ee
This quantity will be essential in our sampling procedure on the
Riemann sphere.

For completeness, let us provide an expression which allows us to translate between
both characterizations of coherent states for integer spin $s=j$. It is
given by the Coherent State (or Bargmann-like) Transform (see e.g.
\cite{CS,Klauder}):
\bea K(\theta,\phi;z)&\equiv& \li
\theta,\phi|z\ld=\sum_{m=-j}^{j}\li \theta,\phi|j,m\ld\li
j,m|z\ld=(1+z\bz)^{-j}\sqrt{\frac{4\pi}{2j+1}}\sum_{m=-j}^{j}
Y^m_j(\theta,\phi) \Upsilon^m_j(z)\nonumber \\ &=&
(1+z\bz)^{-j}\frac{\sqrt{(2j)!}}{2^jj!}(\sin\theta
e^{-i\phi}+2z\cos\theta-z^2\sin\theta e^{i\phi})^j,\label{Bargmann}\eea
which can be seen as a generating function for the spherical functions
$Y^m_j(\theta,\phi)$ when we drop the normalization factor ${\cal N}^{2j}$ from
the last expression.

\section{Sampling Theorem and DFT on $\sd$\label{sampling}}

Sampling techniques consist in the evaluation of a continuous function
(``signal'') on a discrete set of points and later (fully or partially)
recovering the original signal without losing essential information in the
process, and the criteria to that effect are given by various forms of
Sampling Theorems. Basically, the density of sampling points must be high
enough to ensure the reconstruction of the function in arbitrary points
with reasonable accuracy. We shall concentrate on fixed spin holomorphic
(Majorana's) functions and sample them at the roots of unity.

\subsection{Single spin case}

Let us first restrict ourselves to functions in ${\cal H}_s$, i.e., with
well-defined spin or angular momentum $s$. In this case there is a
convenient way to select the sampling points in such a way that
 the resolution operator ${\cal A}$ and/or the reproducing kernel operator ${\cal B}$ are invertible
 and  explicit formulas for their inverses
are available. These are given by the $N^{\rm th}$ roots of unity in the complex
plane, $N\in\mathbb{N}$, which would be associated, by inverse stereographic projection, to
a uniformly distributed set of points in the equator of the Riemann
sphere. The choice of roots of unity is made for convenience, since the
$N^{\rm th}$ roots of any non-zero complex number would also be valid, and
would correspond to different parallels in the Riemann sphere, but then
the formulas obtained are less symmetrical than the ones corresponding to
roots of unity. The most important reason to select roots of unity is that they are associated
with the discrete cyclic subgroup $\mathbb{Z}_N \subset U(1)\subset SU(2)$. The choice $N=2s+1$  corresponds to
\emph{critical sampling}. We shall also discuss the consequences of
\emph{over-sampling}, with $N>2s+1$, and \emph{under-sampling}, with $N<2s+1$, in the following
subsections.

\subsubsection{Over-sampling and critical sampling}

In the case of over-sampling the set ${\cal S}$ generates ${\cal H}_s$,
and the resolution operator ${\cal A}={\cal T}^* {\cal T}$ is
invertible. The case of critical sampling is a particular case of this and
therefore the following discussion also applies to it.

The previous statements are formalized by the following lemma:

 {\lem \label{lemma1} Let ${\cal Q}=\{z_k=e^{2\pi ik/N},\,N\geq 2s+1,\,
k=0,\ldots,N-1\}$ be the discrete subset of the homogeneous space
$Q=SU(2)/U(1)=\mathbb S^2=\bar{\mathbb C}$ made of the $N^{\rm th}$ roots of unity.
The discrete set of CS ${\cal S}=\{|z_k\rangle, z_k\in{\cal Q} \}$
constitutes a discrete frame in ${\cal H}_s$ and the expression
\be
I_{2s+1}=\sum_{k=0}^{N-1}|{z}_k\rangle \langle \tilde{z}_k| =
\sum_{k=0}^{N-1}|\tilde{z}_k\rangle \langle {z}_k |
\label{resolident}\ee
provides a resolution of the identity in ${\cal H}_s$. Here
$|\tilde{z}_k\rangle= {\cal A}^{-1}|z_k\rangle\,,k=0,\ldots,N-1\,,$
denotes the dual frame, and  the resolution operator, ${\cal A}$, is
diagonal in the angular momentum orthonormal basis $B({\cal H}_s)$,
${\cal A}={\rm diag}(\lambda_0,\ldots,\lambda_{2s})$, with
$\lambda_n=\frac{N}{2^{2s}}\tbinom{2s}{n}\,,\,n=0,\ldots,2s$.\label{lemaover}}

%
%

\ni \textbf{Proof.} First, from eq. (\ref{upsilon}) the
expression for the matrix elements of ${\cal T}$ can be obtained,
${\cal T}_{kn}=\langle z_k|s,n-s\rangle =2^{-s}\sqrt{\tbinom{2s}{n}}e^{-i\frac{2\pi
kn}{N}}\,,\,k=0,1,\ldots,N-1\,,\,n=0,1,\ldots,2s$. Then, the resolution
operator turns out to be
\be {\cal A}_{nm}=\sum_{k=0}^{N-1} ({\cal T}_{kn})^*{\cal T}_{km} =2^{-2s}\sqrt{\tbinom{2s}{n}\tbinom{2s}{m}}
\sum_{k=0}^{N-1}e^{2\pi i k(n-m)/N}=N 2^{-2s}\tbinom{2s}{n}\delta_{nm}\,,
\ee
where we have used the well known orthogonality relation
\be \sum_{k=0}^{N-1}\left( e^{2\pi i(n-m)/N}\right)^k=\left\{\ba{l}
N,\;{\rm if}\; n=m \,{\rm mod}\, N\, \\ 0,\;{\rm if}\; n\not=m \,{\rm mod}
\, N\,\ea\right\}=N \delta_{nm}\,,\label{exponencial}\ee
since $N\geq 2s+1$. Therefore ${\cal A}$ is diagonal with non-zero diagonal elements, thus it is
invertible and a dual frame and a (left) pseudoinverse for $\ct$ can be constructed,
${\cal T}_l^+\equiv \ca^{-1}{\cal T}^*$, providing, according to eq. (\ref{resolucionidentidad}),
a resolution of the identity.$\blacksquare$

{\rem It is interesting to rewrite this proof in terms of Rectangular Fourier Matrices (see
Appendix \ref{RFM}). Let
$D={\rm diag}(\lambda_0,\ldots,\lambda_{2s})$ be a diagonal $(2s+1)\times(2s+1)$ matrix,
then $\ct = \cf_{N,2s+1} {\,} D^{1/2} = \cf_N\circ \iota_{N,2s+1}\circ D^{1/2}$. From this
the expression of $\ca=\ct^* \ct=D$ is readily recovered, and also $\cb=\ct \ct^*$ is
seen to be $\cb=\cf_N {\,} D^\uparrow {\,}\cf_N^*$, where
$D^\uparrow =\left( \ba{c|c} D & 0 \\ \hline  0 & 0\ea \right)_{N\times N}$ (see Appendix
\ref{RFM}). Note that $\cb$ is a singular $N\times N$ matrix with only $2s+1$ non-zero eigenvalues
$\lambda_n\,,\,n=0,1,\ldots,2s$, and that they coincide with those of $\ca$. In fact,
$\cb=\cf_N {\,} \ca^\uparrow {\,}\cf_N^*$.}

{\lem \label{lemma2} Under the conditions of the previous lemma, the operator
$P=\ct \ct_l^+= \cf_N{\,} P_{2s+1}{\,} \cf_N^* $ is an orthogonal projector onto a
$(2s+1)$-dimensional subspace of $\mathbb{C}^N$, the range of $\ct$.}

\ni \textbf{Proof.} By direct computation (and using Appendix A),
\bea
P &=&\ct
\ct_l^+=\ct \ca^{-1}\ct^*=\cf_N{\,} \iota_{N,2s+1}{\,} D^{1/2}{\,}
D^{-1}{\,} D^{1/2}{\,} p_{2s+1,N}{\,} \cf_N^*=
\cf_N{\,}\iota_{N,2s+1}{\,} p_{2s+1,N}{\,} \cf_N^* \nn \\ &=& \cf_N{\,}
P_{2s+1}{\,}\cf_N^*\,,
\eea
where $P_{2s+1}=(I_{2s+1})^\uparrow = \left(
\ba{c|c} I_{2s+1} & 0 \\ \hline  0 & 0\ea \right)_{N\times N}$. This
clearly shows that $P$ is an orthogonal projector, unitarily equivalent to
$P_{2s+1}$ and that $P\ct=\ct$.

{\thm \textbf{(Reconstruction formula)} Any function $\psi\in {\cal H}_s$
can be reconstructed from $N\geq 2s+1$ of its samples (the data)
$\Psi(z_k)\equiv \langle z_k|\psi\rangle$, at the sampling points
$z_k=e^{2\pi ik/N},\, k=0,\ldots,N-1$,
by means of
\be
\Psi(z)=\langle
z|\psi\rangle=\sum_{k=0}^{N-1}\Psi(z_k)\Xi(zz_k^{-1}),\label{reconstruccionover} \ee
where
\be
\Xi(z)=\frac{2^s}{N}(1+z\bz)^{-s}\frac{1-\bz^{2s+1}}{1-\bz}  \ee plays the
role of a ``sinc-type function''.

\label{TeoremaReconstruccionOver}}

\ni \textbf{Proof.} {}From the resolution of the identity
(\ref{resolident}), any $\psi\in {\cal H}_s$ can be written as
$|\psi\rangle = \sum_{k=0}^{N-1}\Psi(z_k) |\tilde{z}_k\rangle$, and
therefore $\Psi(z)=\langle z|\psi\rangle= \sum_{k=0}^{N-1}\Psi(z_k)\langle
z|\tilde{z}_k\rangle$. Using that $|\tilde{z}_k\rangle =
\ca^{-1}|z_k\rangle$, we derive that \be \langle z|\tilde{z}_k\rangle=
\frac{1}{\sqrt{N}}\sum_{n=0}^{2s} \lambda^{-1/2}_n e^{2\pi ikn/N} \langle
z|s,n-s\rangle = \frac{2^s}{N}{\cal N}^{2s}\sum_{n=0}^{2s}
(\bar{z}\bar{z}_k^{-1})^n \equiv \Xi(zz_k^{-1})\,,\qquad
k=0,1,\ldots,N-1\,, \ee
where eq. (\ref{upsilon}) has been used.
$\blacksquare$

{\rem \label{lagrange} It is interesting to note that eq.
(\ref{reconstruccionover}) can be interpreted as a Lagrange-type
interpolation formula, where the role of Lagrange polynomials are played
by the functions $L_k(z)=\Xi(zz_k^{-1})$, satisfying the ``orthogonality
relations" $L_k(z_l)=\Xi(z_lz_k^{-1})=P_{lk}$, where $P$ is the projector
of Lemma \ref{lemma2}. In the case of critical sampling, $N=2s+1$, the
usual result $L_k(z_l)=\delta_{lk}$ is recovered, but for the strict
oversampling case, $N>2s+1$, a projector is obtained to account for the
fact that an arbitrary set of overcomplete data
$\Psi(z_k),\,k=0,\ldots,N-1$, can be incompatible with $|\psi\rangle\in
{\cal H}_s$. }

A reconstruction in terms of the Fourier coefficients can be directly obtained
by means of the (left) pseudoinverse of the frame operator $\ct$:

{\cor \label{corolarioover} The Fourier coefficients $a_m$ of the expansion
$|\psi\rangle = \sum_{m=-s}^s a_m|s,m\rangle$  of any $\psi\in{\cal H}_s$
in the angular momentum orthonormal basis $B({\cal H}_s)$ can be
determined in terms of the data $\Psi(z_k)=\langle z_k|\psi\rangle$
as
\be a_{n-s}=\frac{2^s}{N}\tbinom{2s}{n}^{-1/2}
\sum_{k=0}^{N-1}\Psi(z_k) e^{2\pi ikn/N}\,,\,n=0,\ldots,2s\,.
\label{coeffourierover} \ee }

\ni \textbf{Proof.} Taking the scalar product with
$\langle z_k|$ in the expression of
$|\psi\rangle$, we arrive at the over-determined system of equations
\be \sum_{n=0}^{2s} \ct_{kn}a_{n-s}=\Psi(z_k),\;\;\ct_{kn}=\langle
z_k|s,n-s\rangle, \label{sistema} \ee
which can be solved by left multiplying it by the (left) pseudoinverse of
$\ct$, $\ct_l^+=(\ct^*\ct)^{-1}\ct^*=\ca^{-1}\ct^*$. Using the
expressions of $\ca^{-1}={\rm
diag}(\lambda_0^{-1},\lambda_1^{-1},\ldots,\lambda_{2s}^{-1})$, given in
Lemma \ref{lemma1}, and the matrix elements $\ct_{kn}$, given by the
formula (\ref{upsilon}), we arrive at the desired result.$\blacksquare$

{\rem \label{coeFdata} Actually, using vector notation, we have
$\ct\vec{a}=\vec{\Psi}$, where $\vec{a}=(a_{-s},\ldots,a_{s})$, and
$\vec{\Psi}$ denotes the vector of samples $\Psi(z_k)\,,\,k=0,\ldots,N-1$.
Using the (left) pseudoinverse of $\ct$ we can solve it obtaining
$\vec{a}= D^{-1/2}{\,} p_{2s+1,N}{\,} \cf_N^* \vec{\Psi}$, which
coincides with eq. (\ref{coeffourierover}). Note also that the last
expression is a map from $\mathbb{C}^N$ to $\mathbb{C}^{2s+1}\approx {\cal
H}_s$ due to the presence of the projector $p_{2s+1,N}$ (see Appendix
\ref{RFM}), and this prevents the appearance of infinities in the
reciprocal of the binomial coefficient $\tbinom{2s}{n}^{-1/2}$ with
$n>2s$. This is clearer if we apply $\ct$ to the expression of $\vec{a}$
to obtain $\ct \vec{a}= \cf_N{\,} \iota_{N,2s+1}{\,} D^{1/2}{\,}
D^{-1/2}{\,} p_{2s+1,N}{\,} \cf_N^* \vec{\Psi} = P \vec{\Psi}$, that
is, the data $\vec{\Psi}$ should be first projected in order to obtain a
compatible set of data. }

Next we provide an interesting expression.

%
%

{\prop If we define the ``{dual data}'' as $\Gamma(k)\equiv \langle
\tilde{z}_k|\psi\rangle$, then they are related to the data
$\Psi(k)\equiv\Psi(z_k)=\langle z_k|\psi\rangle$ through the convolution
product
\be \Gamma(k)=[\Delta * \Psi](k)=\sum_{l=0}^{N-1}
\Delta(k-l)\Psi(l),\label{wavecoef}\ee
where $\Delta(k)$ (the filter) turns out to be the Rectangular Fourier
Transform of
$\vec\delta\equiv(\lambda_0^{-1},\ldots,\lambda_{2s}^{-1})$, i.e.,
\be
 \Delta(k)=[\cf_{N,2s+1} \delta](k)=\frac{1}{\sqrt{N}}\sum_{n=0}^{2s}\lambda_n^{-1}e^{-i2\pi nk/N}
 =\frac{2^{2s}}{N^{3/2}}\sum_{n=0}^{2s}
 \tbinom{2s}{n}^{-1}e^{-i2\pi nk/N}.\label{filter}\ee
\label{TeoremaReconstruccionWaveOver} } \ni \textbf{Proof.} Applying
(\ref{resolident}) to $\psi$ we obtain:
\[|\psi\rangle=\sum_{k=0}^{N-1}\Gamma(k)|z_k\rangle.\] Taking the scalar product with
$\langle z_l|$ in the last equation,  we arrive at $\vec{\Psi} =
\cb\vec{\Gamma}$, where $\cb=\ct\ct^*$ shows a circulant matrix
structure (see Appendix \ref{Circulantes}). Using the diagonalization
$\cb=\cf_N D^\uparrow \cf_N ^*$ of $\cb$, a Moore-Penrose
pseudoinverse can be computed as $\cb^+=\cf_N (D^{-1})^\uparrow
\cf_N^*$, and this allows us to obtain $\vec{\Gamma} = \cb^+\vec\Psi =
\cf_N (D^{-1})^\uparrow \cf_N^*\vec\Psi$. This last expression, by
duality, can be interpreted as the convolution $\vec\Gamma = \vec\Delta *
\vec\Psi$ between the data and the filter (\ref{filter}). $\blacksquare$

{\rem \label{remarkconvolucion} The relation between $\Psi(k)$ and
$\Gamma(k)$ is simply a ``change of basis'', but with non-orthogonal sets
of generators $\{|z_k\rangle\}$ and $\{|\tilde{z}_k\rangle\}$. Due to the
particular choice of sampling points, the change of basis involves Fourier
transforms, and this can be interpreted as a convolution.}

{\rem For high spin values $s>>1$ (and $N\geq 2s+1$), it is easy to realize that the filter
(\ref{filter}) acquires the simple form
\be
\Delta(k)= \frac{2^{2s}}{N^{3/2}}\left(1+e^{i2\pi r k/N}+O(\frac{1}{2s})\right)\,. \ee
where $r=N-2s$.
There is also a more manoeuvrable closed expression for the exact value of
the filter zero mode
\be
 \Delta(0)=\frac{2^{2s}}{N^{3/2}}\sum_{n=0}^{2s}\tbinom{2s}{n}^{-1}=\frac{2s+1}{N^{3/2}}\sum_{n=0}^{2s}\frac{2^n}{n+1}
 \ee
where we have used the result of the Ref. \cite{Sury} concerning sums of
the reciprocals of binomial coefficients. For large values of $s$ we can
also prove that
\be
\lim_{s\to\infty} \sum_{n=0}^{2s} \tbinom{2s}{n}^{-1}= 2. \ee
 }

In the case of critical sampling all formulae are still valid, we only
have to substitute $N=2s+1$, the difference being that $\ct$ is directly
invertible and $\ct^{-1}=\ct_l^+$. The projector $P$ is the identity, and $\ca$ and $\cb$ are
both invertible. The reason for considering the case of oversampling is twofold: first, by its
intrinsic interest leading to overcomplete frames, and second, in order to apply fast
extensions (as FFT, see \cite{Cooley}) of the reconstruction algorithms it would be useful to consider
$N$ the smallest
power of 2 greater or equal to $2s+1$.

\subsubsection{Under-sampling and critical sampling}
\label{undersampling}

Let us suppose now that the number of sampling points is $N\leq 2s+1$. We
shall see that, for $N< 2s+1$, we cannot reconstruct exactly an arbitrary
function $\psi\in {\cal H}_s$ but its orthogonal projection
$\hat{\psi}\equiv P_N\psi$ onto the subspace $\hat{\cal H}_s$ of ${\cal
H}_s$ spanned by the discrete set ${\cal S}=\{|z_k\rangle, k=0,\dots,N\}$
of CS. In other words, the restriction to this discrete subset implies a
loss of information.

This loss of information translates to the fact that the resolution
operator $\ca$  is not invertible and therefore we do not have a frame nor
a resolution of the identity like in the previous subsection, see the
discussion  at the end of Sec. \ref{CSandFrames}. But, since the set
${\cal S}$ is linearly independent, we can construct another operator, the
overlapping kernel ${\cal B}=\ct \ct^*$, which is invertible and
provides a partial reconstruction formula. In addition, the overlapping
kernel operator has a circulant structure, and this provides a deep
insight in the reconstruction process.

Let us formalize again the previous assertions.

{\lem \label{lemmaunder1} Let ${\cal Q}=\{z_k=e^{2\pi ik/N},\,\,
k=0,\ldots,N-1\}$ the discrete subset of the homogeneous space
$Q=SU(2)/U(1)=\mathbb S^2=\bar{\mathbb C}$ made of the $N^{\rm th}$ ($N\leq 2s+1$) roots
of unity. The pseudo-frame operator $\ct:{\cal H}_s\to \mathbb C^N$ given
by $\ct(\psi)=\{\langle z_k|\psi\rangle, z_k\in {\cal Q}\}$ [remember the
construction after Eq. (\ref{pbiop2})] is such that the overlapping kernel
operator ${\cal B}=\ct\ct^*$ is an $N\times N$ Hermitian positive
definite invertible matrix, admitting  the eigen-decomposition
${\cal B} ={\cal F}_N\hat{D}{\cal F}_N^*$, where $\hat{D} ={\rm
diag}(\hat\lambda_0,\ldots,\hat\lambda_{N-1})$ is a diagonal matrix with
$\hat\lambda_k=\frac{N}{2^{2s}}\sum_{j=0}^{\bar{q}-1}\tbinom{2s}{k+jN}$,
 $\bar{q}$ being the ceiling of $(2s+1)/N$.
}

\ni \textbf{Proof.}
Let us see that the eigenvalues $\hat\lambda_k$ of ${\cal B}=\ct\ct^*$
are indeed all strictly positive and hence ${\cal B}$ is invertible. This can be done
using RFM or taking advantage of the circulant structure of ${\cal B}$ (see Appendix
\ref{Circulantes}). With RFM we start with the expression of
$\ct=\cf_{N,2s+1}{\,} D^{1/2}$ to obtain
$\cb=\ct\ct^* = \cf_{N,2s+1}{\,} D{\,} \cf_{N,2s+1}^*$, which should be further
worked on in order to fully diagonalize it.

This can be done by using the ``trick" mentioned in Appendix \ref{RFM} consisting in
enlarging the RFM $\cf_{N,2s+1}$ to  $\cf_{N\bar{M}}$ where $\bar{M}$ is the smaller
multiple of $N$ greater or equal to $2s+1$, and $\bar{q}=\bar{M}/N$ is the ceiling of
$(2s+1)/N$
(see Appendix \ref{RFM}). In this way $\cf_{N\bar{M}}$ always contains an integer number
of ordinary Fourier matrices $\cf_N$.

Using this we obtain that $\cb= \cf_{N\bar{M}}{\,} D^\uparrow{\,} \cf_{N\bar{M}}^*$
where $D^\uparrow$ is the extension of $D$ to a $\bar{M}\times\bar{M}$ matrix, and
with a little of algebra  the expression
$\cb= \cf_{N}{\,} \hat{D}{\,} \cf_{N}^*$ is obtained, where $\hat{D} ={\rm
diag}(\hat\lambda_0,\ldots,\hat\lambda_{N-1})$ and
\be
\hat\lambda_k=\sum_{l=0}^{\bar{q}-1}\lambda_{k+lN}=\frac{N}{2^{2s}}\sum_{l=0}^{\bar{q}-1}\tbinom{2s}{k+lN}
\label{eigenBunder} \ee
All the eigenvalues are strictly positive and therefore $\cb$ is invertible.
$\blacksquare$

Following Sec. 2, we introduce the following result:

{\lem \label{lemmaunder2} Under the conditions of the previous Lemma, the
set $\{|\tilde{z}_k\rangle =\sum_{k=0}^{N-1}
(\cb^{-1})_{lk}|z_l\rangle\,,k=0,\ldots,N-1\}$ constitutes a dual
pseudo-frame for ${\cal S}$, the operator $P_{\cal S}=\ct_r^+\ct$ is an
orthogonal projector onto the subspace of ${\cal H}_s$ spanned by ${\cal
S}$, where $\ct_r^+=\ct^* \cb^{-1}$ is a (right) pseudoinverse for
$\ct$, and
\be
 \sum_{k=0}^{N-1} |\tilde{z}_k\rangle\langle z_k| =
\sum_{k=0}^{N-1} |z_k\rangle\langle \tilde{z}_k| =P_{\cal S}\label{resolproy}
\ee
provides a resolution of the projector $P_{\cal S}$.}

\ni \textbf{Proof.} If we define $\ct_r^+=\ct^* \cb^{-1}$ it is easy to check that
$\ct \ct_r^{+}=I_N$ is the identity in $\mathbb{C}^N$. In the same way,
$P_{\cal S}=\ct_r^+\ct$ is a projector since
$P_{\cal S}^2=\ct_r^+\ct\ct_r^+\ct=\ct_r^+\ct=P_{\cal S}$ and it is orthogonal
$P_{\cal S}^*=(\ct^* \cb^{-1}\ct)^* = \ct^* \cb^{-1}\ct =P_{\cal S}$ since $\cb$
is self-adjoint. The resolution of the projector is provided by eq. (\ref{resolucionproyector}).
$\blacksquare$


Although the full reconstruction of the original signal is not possible in the case of
undersampling, a partial reconstruction is still possible in the following sense.

{\thm \textbf{(Partial reconstruction formula)} Any function $\psi\in
{\cal H}_s$ can be partially reconstructed from $N\leq 2s+1$ of its
samples (the data) $\Psi(z_k)\equiv \langle z_k|\psi\rangle$, at the
sampling points $z_k=e^{2\pi ik/N},\, k=0,\ldots,N-1$, by the alias
$|\hat{\psi}\rangle=P_{\cal S}|\psi\rangle$, by means of
\be
\hat{\Psi}(z)=\langle
z|\hat{\psi}\rangle=\sum_{k=0}^{N-1}\Psi(z_k)\hat{\Xi}(zz_k^{-1}),\label{reconstruccionunder} \ee
where \be \hat{\Xi}(z)=\frac{2^s}{N}(1+z\bz)^{-s} \sum_{p=0}^{N-1}
\hat{\lambda}^{-1}_p \sum_{l=0}^{\bar{q}-1} \lambda_{p+lN} \bar{z}^{p+lN}
\ee
plays the role of a ``sinc-type function''.
\label{TeoremaReconstruccionUnder}}

\ni \textbf{Proof.} The proof follows the same lines as in Theorem
\ref{TeoremaReconstruccionOver}. {}From the resolution of the projector
(\ref{resolproy}), any $\psi\in {\cal H}_s$ has a unique alias
$\hat{\psi}=P_{\cal S}\psi$ which can be written as $|\hat{\psi}\rangle =
\sum_{k=0}^{N-1}\Psi(z_k) |\tilde{z}_k\rangle$, and therefore
$\hat{\Psi}(z)=\langle z|\hat{\psi}\rangle=
\sum_{k=0}^{N-1}\Psi(z_k)\langle z|\tilde{z}_k\rangle$. Using that
$|\tilde{z}_k\rangle = \sum_{l=0}^{N-1}(\cb^{-1})_{lk}|z_l\rangle$, we
derive that \be \langle z|\tilde{z}_k\rangle=
\frac{1}{\sqrt{N}}\sum_{l=0}^{N-1}(\cb^{-1})_{lk} \sum_{n=0}^{2s}
\lambda_n^{1/2} e^{\frac{2\pi i kn}{N}} \langle z|s,n-s\rangle =
\frac{2^s}{N} {\cal N}^{2s} \sum_{p=0}^{N-1} \hat{\lambda}_p^{-1}
\sum_{l=0}^{\bar{q}-1} \lambda_{p+lN} (\bar{z}\bar{z}_k^{-1})^{p+lN} =
\hat{\Xi}(zz_k^{-1}) \ee
where eq. (\ref{upsilon}) and the orthogonality relations (\ref{exponencial}) have been used
(but with $N\leq 2s+1$, as in Appendix \ref{Circulantes}).
$\blacksquare$

{\rem \label{lagrangeunder} As in the case of oversampling, eq.
(\ref{reconstruccionunder}) can be interpreted as a Lagrange-type
interpolation formula, where the role of Lagrange polynomials are played
by the functions $\hat{L}_k(z)=\hat{\Xi}(zz_k^{-1})$, this time satisfying
the proper orthogonality relations
$\hat{L}_k(z_l)=\hat{\Xi}(z_lz_k^{-1})=\delta_{lk}$. The reason for this
is that in the case of under-sampling there is not an overcomplete set of
data, and therefore the ``Lagrange functions" are orthogonal, although not
complete. }


A partial reconstruction can also be obtained, in a natural way, from the ``dual data":

{\prop If we define the ``{dual data}'' as $\Gamma(k)\equiv \langle
\tilde{z}_k|\psi\rangle$, then they are related to the data $\Psi(k)=\langle
z_k|\psi\rangle$ through the convolution product
\be \Gamma(k)=[\hat{\Delta} * \Psi](k)=\sum_{l=0}^{N-1}
\hat{\Delta}(k-l)\Psi(l),\label{wavecoefunder}\ee
where $\hat{\Delta}(k)$ (the filter) turns out to be the discrete Fourier transform of
$\hat\delta\equiv(\hat\lambda_0^{-1},\ldots,\hat\lambda_{N-1}^{-1})$, where $\hat\lambda_k$ are the
eigenvalues (\ref{eigenBunder}) of the overlapping kernel operator
$\cb$:
\be
 \hat\Delta(k)=[\cf_N \hat\delta](k)=\frac{1}{\sqrt{N}}\sum_{n=0}^{N-1}\hat\lambda_k^{-1}e^{-i2\pi nk/N}
 \label{filterunder}.\ee
\label{TeoremaReconstruccionWaveUnder} }

\ni \textbf{Proof.} The proof follows the same lines as in Proposition
\ref{TeoremaReconstruccionWaveOver}, with the difference that now $\cb$ is
not singular and there is no need for a pseudoinverse. From $\vec\Psi =
\cb \vec{\Gamma}$ and using
 the diagonalization $\cb=\cf_N \hat{D}\cf_N ^*$ of $\cb$, the inverse is
 directly $\cb^{-1}=\cf_N \hat{D}^{-1} \cf_N^*$,
and this allows to obtain $\vec{\Gamma} = \cb^{-1}\vec\Psi = \cf_N
\hat{D}^{-1} \cf_N^*\vec\Psi$. This last expression, by duality, can be
interpreted as the convolution
$\vec{\Gamma} = \vec{\hat{\Delta}} * \vec\Psi$ between the data and the filter
(\ref{filterunder}).$\blacksquare$

The comments made in Remark \ref{remarkconvolucion} also apply here.

Again, a reconstruction in terms of the Fourier coefficients can be
directly obtained by means of the (right) pseudoinverse of the frame
operator $\ct$:

{\cor \label{corolariounder} The Fourier coefficients $\hat{a}_m$ of the
expansion $|\hat{\psi}\rangle = \sum_{m=-s}^s \hat {a}_m|s,m\rangle$ of
the alias of any $\psi\in{\cal H}_s$ in the angular momentum orthonormal
basis $B({\cal H}_s)$ can be determined in terms of the the data
$\Psi(k)=\langle z_k|\psi\rangle$ as
\be
\hat{a}_{n-s}=\frac{N}{2^s}\tbinom{2s}{n}^{1/2} \sum_{k=0}^{N-1}e^{2\pi
ikn/N} \sum_{l=0}^{N-1} (\cb^{-1})_{kl}\Psi(l)\,,\,n=0,\ldots,2s\,.
\label{coeffourierunder} \ee }

\ni \textbf{Proof.} Taking the scalar product with
$\langle z_k|$ in the expression of
$|\hat{\psi}\rangle$, we arrive at the  system of equations
\be \sum_{n=0}^{2s} \ct_{kn}\hat{a}_{n-s}=\Psi(k),\;\;\ct_{kn}=\langle
z_k|s,n-s\rangle, \label{sistemaunder} \ee
which can be solved by left multiplying it by the (right) pseudoinverse
of $\ct$, $\ct_r^+=\ct^* =\ct^*\cb^{-1}$. Using the expressions of
$\cb$, given in Lemma \ref{lemmaunder1}, and the matrix elements
$\ct_{kn}$, given by the formula (\ref{upsilon}), we arrive at the desired
result by noting that $\ct_r^+\ct= P_{\cal S}$ and this acts as the
identity on $\hat{a}_{n-s}$. $\blacksquare$

{\rem Using vector notation this can be written as
$\vec{\hat{a}}=\ct_r^+\ct \vec{a}= \ct_r^+\vec{\Psi} = \ct^*
\cb^{-1}\vec{\Psi}$, and this is even simpler in terms of the dual data,
$\vec{\hat{a}}=\ct^* \vec{\Gamma}=D^{1/2}{\,} \cf_{N,2s+1}^*
\vec{\Gamma}$. \label{remarkunder}}

It is interesting to establish the connection between our results and
others in the literature \cite{GMP}.

{\cor \textbf{(Covariant interpolation)} For $0\leq k\leq N-1$ define on
$Q$ the functions $\Phi_k(z)\equiv\langle z|z_k\rangle,\,z\in {\mathbb C}$. Let
$\zeta_0,\dots,\zeta_{N-1}$ be $N$ complex numbers and $\cb_{kl}$ the overlapping kernel
operator. Define on $Q$ the function
\bea \Phi(z)&=&\Phi(z_0,\dots,z_{N-1};\zeta_0,\dots,\zeta_{N-1};z)\nn\\
&\equiv&-\frac{1}{\det(\cb)}\det\left(\ba{llll} 0 & \Phi_1(z) & \dots &
\Phi_{N-1}(z) \\ \zeta_0 & \cb_{0,0} & \dots & \cb_{0,N-1}\\ \vdots
&\vdots &\ddots & \vdots\\ \zeta_{N-1} & \cb_{N-1,0} & \dots &
\cb_{N-1,N-1} \ea\right).\eea Then we have that
\begin{enumerate}
\item $\Phi(z)=\langle z|\phi\rangle$ for some $\phi\in {\cal H}_s$
\item $\Phi$ is a solution of the interpolation problem, i.e.,
$\Phi(z_k)=\zeta_k,\, z=0,\dots,N-1$.
\item $\Phi$ is of minimal norm, in the sense that if $\tilde{\Phi}$ is
any other function on $Q$ with $\tilde{\Phi}(z)=\langle
z|\tilde{\phi}\rangle$, for some $\tilde{\phi}\in {\cal H}_s$, and
$\tilde{\Phi}(z_k)=\zeta_k$, then $||\tilde{\Phi}||\geq ||\Phi||$.
\item The interpolation procedure is invariant under left multiplication in $G$, in the sense that
$U(g)\cb U(g)^*=\cb$ and
\[\Phi(gz_0,\dots,gz_{N-1};\zeta_0,\dots,\zeta_{N-1};gz)=\Phi(z_0,\dots,z_{N-1};\zeta_0,\dots,\zeta_{N-1};z),\]
($gz$ denotes the natural action of the group $G$ on its homogeneous space
$Q=G/H$) so that the left-displaced interpolation problem
$\check{\Phi}(gz_k)=\zeta_k$ is solved by the function $\check{\Phi}(z)=\Phi(g^{-1}z)$.
\end{enumerate}
}
\ni \textbf{Proof.} It is a direct consequence of Theorem
\ref{TeoremaReconstruccionUnder} if we identify the data
$\zeta_k=\Psi(z_k),\,\Phi(z)=\hat{\Psi}(z)$ and
$\sum_{k=0}^{N-1}(\cb^{-1})_{kl}\Phi_l(z)=\hat{\Xi}(zz_k^{-1})$. The fact
that $\Phi$ is of minimal norm is a direct consequence of the
orthogonality of the projector $P_{\cal S}$. The invariance under left
multiplication is a consequence of the invariance of the overlapping
kernel $\cb$ under left multiplication.$\blacksquare$

\subsection{Several spin case}

The case of several spins, i.e., band limited functions, is more involved than the
single spin case, and it is not so easy to select the sampling points in such a way that
an explicit expression for the inverse of the resolution or overlapping kernel operators be available.

Denoting by\footnote{For the time being we shall restrict ourselves to integer values
of spin, in order to compare with standard Fourier analysis on the sphere.}
${\cal H}^{(J)}=\bigoplus_{s=0}^J {\cal H}_s$ the Hilbert space
of band-limited functions, up to spin $J$, the set of coherent states can be defined
in an analogous way to the single spin case.

First, let us denote by  $U^{(J)}(z,\bar{z})=\bigoplus_{s=0}^J U_s(z,\bar{z})$ the
unitary and reducible representation of $SU(2)$ acting on ${\cal H}^{(J)}$, where
$U_s(z,\bar{z})$ stands for the unitary and irreducible representation of spin $s$.
The Hilbert space ${\cal H}^{(J)}$ has an orthogonal basis given by $\{|s,m\rangle\}$,
in such a way that
$I_{{\cal H}^{(J)}}= \frac{1}{J+1} \sum_{s=0}^J\sum_{m=-s}^s |s,m\rangle\langle s,m|$ is a resolution
of the identity. Selecting the fiducial vector
$|\gamma\rangle^J=\frac{1}{\sqrt{J+1}}\bigoplus_{s=0}^J|s,s>$, the set of coherent
states is defined as $|z\rangle^J=U^{(J)}(z,\bar{z})|\gamma\rangle^J$.

The CS overlap, for the several spins case, is now
\be C^{(J)}(z,z')=\langle z|z'\rangle^J=\frac{1}{J+1}\sum_{s=0}^J
\frac{(1+z'\bar{z})^{2s}}{(1+z\bar{z})^s(1+z'\bar{z}')^s}.\label{CSoverlap2}
\ee

The first, naive choice, of sampling points would be the $N^{\rm th}$ roots of
unity, $z_k=e^{2\pi i k/N}$, where now $N={\rm dim}{\cal H}^{(J)}=(J+1)^2$
in order to have critical sampling. In this way the operators ${\cal
A}^{(J)}$ and $\cb^{(J)}$ would have nice structure and their inverse
matrices would be easily computed.

However, the following negative result prevents us from proceeding in this way:

{\prop For $N\geq 2J+1$, the overlapping kernel operator ${\cal B}^{(J)}$ has rank $2J+1$.}
%
\ni \textbf{Proof.} Let $\lambda^{s}$, $\ct^{s}$, $\cb^s$ and $D_{2s+1}$ the eigenvalues,
frame, overlapping kernel operators  and diagonal matrix appearing in the previous sections corresponding to angular momentum $s$. Then
the frame operator $\ct^{(J)}:{\cal H}^{(J)}\rightarrow \mathbb{C}^N$ can be written as
a $N\times (J+1)^2$ matrix given by
$\ct^{(J)}_{k,(s,n)}=\ct^s_{kn}=\cf_{N,2s+1}\,D_{2s+1}^{1/2}$.

Then $\cb^{(J)}= \ct^{(J)}\ct^{(J)*}=\cb^0+\cb^1+\cdots + \cb^J=\cf_N\,(D_{1}^\uparrow +
D_3^\uparrow + \cdots + D_{2J+1}^\uparrow)\,\cf_N^* = \cf_N
\,\tilde{D}_{2J+1}^\uparrow\,\cf_N^*$,
where $\tilde{D}_{2J+1}$ is a diagonal matrix with eigenvalues
\be \tilde{\lambda}_n=\frac{1}{J+1}\sum_{s=\overline{({n-1})/{2}}}^J
\lambda_n^s\,,\qquad n=0,1,\ldots 2J \ee
where $\overline{({n-1})/{2}}$ stands for the ceiling of
$\frac{n-1}{2}$.$\blacksquare$

The proof could also have been done using the circulant structure of $\cb^{(J)}$, which
can be written as $\cb^{(J)}_{kl}=\cc_{l-k}$, where now
\[\cc_k\equiv
\frac{1}{J+1} \sum_{s=0}^J \frac{1}{2^{2s}}\left(1+e^{2\pi ik/N}\right)^{2s},\]
and computing its eigenvalues as in Appendix \ref{Circulantes}.

Therefore, putting all sampling points in the equator of the Riemann sphere is not a good
choice, and other alternatives should be looked for. The problem is that other choices
of sampling points lead to resolution operators with less structure and therefore
without the possibility of having an explicit inverse.

Another possibility is to use an equiangular grid in $(\theta,\phi)$, as the one used
in \cite{DH}. If
$(\theta_j,\phi_k)=(\frac{\pi}{N}j,\frac{2\pi}{N}k)\,,j,k=0,1,\ldots,N-1$
is a  grid of $N^2$ points in the sphere, where $N\geq J+1$, the
corresponding points in the complex plane by stereographic projection are
given by
$z_j^k=e^{i\phi_k}\tan\frac{\theta_j}{2}=e^{i\frac{2\pi}{N}k}\tan(\frac{\pi}{2N}j)=r_je^{i\frac{2\pi}{N}k}$.
However, it can be checked that in this case the resolution operator
is also singular.

We shall follow a mixture of both approaches, consisting in using as sampling points
the $(2s+1)^{\rm th}$ roots of $(r_s)^{2s+1}$, for $s=0,1,\ldots,J$. Here $r_s$ is a positive number
depending on $s$, in such a way that if $s\neq s'$ then $r_{s}\neq r_{s'}$. Thus, we shall
continue to use $N=(J+1)^2$ sampling points but distributed in circles of different
radius. In the
Riemann sphere, these would be distributed in different parallels, one for each value of
spin. These points are given by
$$z_m^{(s)}=r_s\;e^{\frac{2\pi i m}{2s+1}}\,,\qquad s=0,\ldots,J\,,\quad m=0,\ldots,2s\,, $$
where $s$ denotes spin index and $m$ the index for the roots.

The frame operator $\ct$ is a $(J+1)^2\times (J+1)^2$ square matrix with a block structure given
by $(2s'+1)\times (2s+1)$ blocks
\be
\ct^{s',s}=\cf_{2s'+1,2s+1}(D_{2s+1}^{s',s})^{1/2}\,,
\quad\hbox{with}\quad D_{2s+1}^{s',s}={\rm diag}(\lambda^{s',s}_0,\ldots,\lambda^{s',s}_{2s})
\ee
where $\lambda^{s',s}_n=\frac{1}{J+1}\frac{2s'+1}{(1+r_{s'}^2)^{2s}}\tbinom{2s}{n} r^{2n}_{s'}$. The diagonal blocks
are the frame operators for the case of critical sampling with fixed spin $s$, for each value of
$s=0,1,\ldots,J$, and they coincide with the previous expressions fixing $r_s=1$ (up to the factor
$\frac{1}{J+1}$).

The resolution operator $\ca$ and the overlapping kernel operator $\cb$ share this block structure,
with blocks given by
\bea
\ca^{s',s} &= & \sum_{s''=0}^J (\ct^{s'',s'})^*\ct^{s'',s}   = \sum_{s''=0}^J ( D^{s'',s'}_{2s'+1})^{1/2}\cf_{2s''+1,2s'+1}^*\cf_{2s''+1,2s+1}( D^{s'',s}_{2s+1})^{1/2}\nn\\
\cb^{s',s} &= & \sum_{s''=0}^J \ct^{s',s''}(\ct^{s,s''})^* = \sum_{s''=0} \cf_{2s'+1,2s''+1}
( D^{s',s''}_{2s''+1})^{1/2}( D^{s,s''}_{2s''+1})^{1/2}\cf_{2s+1,2s''+1}^*
\eea

The overlapping kernel operator $\cb$ can be computed directly from the CS overlap (\ref{CSoverlap2}) evaluated a the
sampling points, turning out to be
\bea {\cal B}^{a,b}_{m,n}&\equiv&  \langle
z_{m}^{(a)}|z_{n}^{(b)}\rangle^J=\frac{1}{J+1}\sum_{s=0}^J\left(\frac{\left(1+r_ar_be^{2\pi
i\frac{n(2a+1)-m(2b+1)}{(2a+1)(2b+1)}}\right)^2}{(1+r_a^2)(1+r_b^2)}\right)^s\nonumber\\
&=&\left\{\ba{l}
1,\;{\rm if}\;z^{(a)}_m=z^{(b)}_n\\\\
\frac{1}{J+1}\frac{1-(\kappa^{a,b}_{m,n})^{J+1}}{1-\kappa^{a,b}_{m,n}},\;{\rm
otherwise}\ea\right.\eea

where \be \kappa^{a,b}_{m,n}\equiv\frac{(1+r_ar_be^{2\pi
i\frac{n(2a+1)-m(2b+1)}{(2a+1)(2b+1)}})^2}{(1+r_a^2)(1+r_b^2)}\ee
is the multiplier of a geometric sum.

The overlapping kernel operator ${\cal B}^{a,b}_{m,n}$
is an Hermitian matrix having the following structure:

$$ {\footnotesize
\left(\begin{array}{c|c|c|c|c|c|c}
    {\rm circ}(1) & B_{01} & B_{02} & \ldots & B_{0k} & B_{0\;k+1} & \ldots \\
  \hline B_{01}^* & {\rm circ}(\cc_ 0^{(1)},\cc_1^{(1)},\cc_2^{(1)}) & B_{12} & \ldots & B_{1k} & B_{1\;k+1} & \ldots \\
  \hline B_{02}^* & B_{12}^* & {\rm circ}(\cc_ 0^{(2)},\ldots,\cc_4^{(2)}) & \ldots & B_{2k} &
  B_{2\;k+1} & \ldots \\
  \hline\vdots & \vdots & \vdots & \ddots & \vdots & \vdots & \vdots \\
  \hline B_{0k}^* & B_{1k}^* & B_{2k}^* & \ldots & {\rm circ}(\cc_ 0^{(k)},\ldots,\cc_{2k}^{(k)}) & B_{k\;k+1} & \ldots \\
  \hline B_{0\;k+1}^* & B_{1\;k+1}^* & B_{2\;k+1}^* & \ldots & B_{k\;k+1}^* & \ddots & \ldots \\
  \hline\vdots & \vdots & \vdots & \vdots & \vdots & \vdots & \ddots \\
\end{array}\right)%
}$$
\newline

The diagonal blocks are circulant matrices of dimension $2s+1$, with
${\cal C}_n^{(s)}= \displaystyle{ \frac{(1+r_s^2 e^{2\pi i
n/(2s+1)})^2}{(1+r_s^2)^2}}$, and the non-diagonal blocks $B_{pq}$ are
matrices of dimension $(2p+1)\times(2q+1)$.

The overlapping kernel operator ${\cal B}^{a,b}_{m,n}$ is not a circulant
matrix\footnote{This is traced back to the fact that the sampling points
do not form an Abelian group. Only the sets of the form
$z_m^{(s)}\,,m=0,1,\ldots,2s$, with fixed $s$ form cyclic subgroups, and
they are responsible for the appearance of circulant blocks at the
diagonal.}, not even a block circulant matrix, therefore the computation
of its inverse needed for the reconstruction formula must be done
numerically. Even the checking that it is non-singular must be done
numerically.

For different choices of $r_s$, we have checked that the overlapping kernel
operator is invertible, and therefore the reconstruction formula can be
used, although nice expressions like (\ref{reconstruccionover}) or
(\ref{reconstruccionunder}) are not available.

Once the overlapping kernel operator has been inverted, the Fourier coefficients of the signal
can be obtained in the same fashion as in Corollary \ref{corolarioover} or \ref{corolariounder},
using the vector notation  of Remarks \ref{coeFdata} and \ref{remarkunder}.

{\cor \label{corolarioseveral}  The Fourier coefficients $a_m^s$ of the
expansion $|\psi\rangle = \sum_{s=0}^J\sum_{m=-s}^s a_m^s|s,m\rangle$ of
any $\psi\in{\cal H}^{(J)}$ in the angular momentum orthonormal basis
$B({\cal H}^{(J)})$ can be determined in terms of the the data
$\Psi(z_k^{(s)})=\langle z_k^{(s)}|\psi\rangle$ as
\be
\vec{a}=\ct^*\cb^{-1}\vec{\Psi}\,,
\ee
}

In this expression $\vec{a}$ and $\vec{\Psi}$ stand for  vectors formed by
gathering the vectors $\vec{a}$ and $\vec{\Psi}$ of Remarks \ref{coeFdata}
and \ref{remarkunder} for each spin $s$.

\ni \textbf{Proof.} As in corollaries \ref{corolarioover} and
\ref{corolariounder}, using vector notation, we have $\ct \vec{a} =
\vec{\Psi}$. Applying the right pseudoinverse for $\ct$, the desired
result is obtained.$\blacksquare$

{\rem For this choice of sampling points, both $\ca$ and $\ct$ turn out to
be invertible, therefore we could have used the left pseudoinverse of
$\ct$ for the reconstruction, which requires the inverse of $\ca$, o we
could have directly inverted $\ct$. \label{remarkseveral}}

>From the computational point of view, the most expensive step is the inversion of $\cb$ (or $\ca$ or $\ct$),
which is of the order $O(N^3)$, with $N=(J+1)^2$. But this is done only once, and can be stored for future uses.
The determination of the Fourier coefficients from the data requires $O(N^2)$ operations. More efficient algorithms, to
compete with the $O(N\log(N)^2)$ of  \cite{DH} would require taking advantage of the block structure of the
matrices $\cb$ or $\ca$, or maybe choosing a different set of sampling points that leads to more structured matrices,
in such a way that the inverse is easily computed.




\section{Connection with the Euler angle picture}\label{comments}

We have provided reconstruction formulas for Majorana functions $\Psi(z)$
from $N$ of its samples $\Psi_k=\langle z_k|\psi\rangle$ at the sampling
points $z_k=e^{2\pi i k/N}$ in the Riemann sphere. As stated in the
Introduction, the advantage of using this ``complex holomorphic picture'',
instead of the standard ``Euler angle picture'', is twofold: firstly we
can take advantage of the either diagonal or circulant structure of
resolution and overlapping kernel operators, respectively, to provide
explicit inversion formulas and, secondly, we can extend the sampling
procedure to half-integer angular momenta $s$, which could be useful when
studying, for example, discrete frames for coherent states of spinning
particles in Quantum Mechanics.

Moreover, for integer angular momenta $s=j$, we could always pass from one
picture to another through the Bargmann transform (\ref{Bargmann}).
Indeed, let us work for simplicity in the critical case $N=2j+1$ and let
us denote by $\Phi_k=\langle \theta_0,\phi_k|\psi\rangle$ the samples of
the function (\ref{Wigner}), in the Euler angle characterization, at the
sampling points $\theta_0\neq 0,\pi$ and $\phi_k=-\frac{2\pi}{N}k, \,
k=0,\dots, N-1$ (i.e., a uniformly distributed set of $N$ points in a
parallel of the sphere $\mathbb S^2$, but counted clockwise). Denoting by
\be {\cal K}_{kl}\equiv K(\theta_0,\phi_k;z_l)=\langle
\theta_0,\phi_k|z_l\rangle=
\frac{\sqrt{(2j)!}}{2^{2j}j!}e^{i\frac{2\pi}{N}jk}\sin^j(\theta_0)
(1+2\cot(\theta_0)e^{i\frac{2\pi}{N}(l-k)}-e^{i\frac{4\pi}{N}(l-k)})^j\label{discreteBargmann}\ee
a discrete $N\times N$ matrix version of the Bargmann transform
(\ref{Bargmann}), and inserting the resolution of the identity
(\ref{resolident}) in $\langle \theta_0,\phi_k|\psi\rangle$, we easily
arrive to the following expression:
\be \Phi_k=\sum_{l,m=0}^{N-1}{\cal
K}_{kl}\cb^{-1}_{lm}\Psi_m,\label{datapass}\ee which relates data between
both characterizations or pictures through the CS transform and CS overlap
matrices ${\cal K}$ and $\cb$ in (\ref{discreteBargmann}) and
(\ref{resolentriesunder}), respectively.

Except for some values of $\theta_0$ (see later in this section), the
transformation (\ref{datapass}) is invertible and explicit formulas of
${\cal K}^{-1}$ are available. Actually,
${\cal K}$ can be written as the product
$ {\cal K}=\Lambda {\cal Q}$,
\be\Lambda_{kp}=\frac{\sqrt{(2j)!}}{2^{2j}j!}e^{i\frac{2\pi}{N}jk}\sin^j(\theta_0)\delta_{kp},
\;\;{\cal
Q}_{pl}=(1+2\cot(\theta_0)e^{i\frac{2\pi}{N}(l-p)}-e^{i\frac{4\pi}{N}(l-p)})^j\equiv
q_{l-p},\ee of a diagonal matrix $\Lambda$ times a circulant matrix ${\cal
Q}$, which can be easily inverted (following the procedure of Appendix
\ref{Circulantes}) as ${\cal Q}^{-1}={\cal F}_N \Omega^{-1} {\cal
F}_N^*$, where $\Omega={\rm diag}(\omega_0,\dots,\omega_{N-1})$, with
eigenvalues
\be \omega_k=\sum_{n=0}^{N-1}q_n e^{-i\frac{2\pi}{N}k
n}=N\sum_{p=0}^j\sum_{r=0}^p{}'
(-1)^{r}\tbinom{p}{r}(2\cot(\theta_0))^{p-r},\ee
where the prime over $\sum$ implies the restriction $p+r=k$. Therefore, we
can also obtain data in the holomorphic characterization,
$\vec{\Psi}$, from data in the Euler angle characterization, $\vec{\Phi}$,
through the formula
\be \vec{\Psi}=\cb {\cal
Q}^{-1}\Lambda^{-1}\vec{\Phi}={\cal F}_N D \Omega^{-1} {\cal F}_N^*
\Lambda^{-1}\vec{\Phi}\ee which can be seen as a convolution
$\vec{\Psi}=\vec{\Theta}*\vec{\Phi}'$ of the re-scaled data
$\vec{\Phi}'=\Lambda^{-1}\vec{\Phi}$ times the filter $\vec{\Theta}={\cal
F}_N \vec{\theta}$, with ${\theta}_k=\lambda_k/\omega_k$ the quotient of
eigenvalues of $\cb$ and ${\cal Q}$.

Note that there are values of $\theta_0$ for which ${\cal K}$ is not
invertible. Such is the case of
$\theta_0=\pi/2$ (the equator), for which $\omega_k=N(-1)^{k/2}\tbinom{j}{k/2}$ if
$k$ even, and zero otherwise. Let us show that this situation is linked to the fact that
general functions (\ref{Wigner}) in the Euler angle picture can not be reconstructed from its
samples $\Phi_k$ on a uniformly distributed set of $N$ points in the
equator of the sphere. Indeed, let us insert this time the resolution of
unity
$I_{2j+1}=\sum_{m=-j}^{j}|j,m\rangle\langle j,m|$ in  $\langle
\theta_0,\phi_k|\psi\rangle$, with $|\psi\rangle=\sum_{m=-j}^{j}
a_{m}|j,m\rangle$, which results in
\be \Phi_k=\sum_{m=-j}^{j}
\sqrt{\frac{4\pi}{2j+1}}Y_j^{m}(\theta_0,\phi_k)\, a_m,\label{variant}\ee
where we have used the definition (\ref{spherical}). Denoting ${\cal
Y}_{kn}({\theta_0})\equiv\sqrt{\frac{4\pi}{N}}Y_j^{n-j}(\theta_0,\phi_k)$
and knowing from Remark \ref{coeFdata} that the Fourier coefficients
$a_{n-j}$ are given in terms of data $\Psi_k$ trough $\vec{a}=D^{-1/2}
\cf_N^*\vec{\Psi}$, we arrive to a variant of the formula
(\ref{datapass}):
 \be
 \vec{\Phi}={\cal Y}({\theta_0}) D^{-1/2} \cf_N^* \vec{\Psi},\ee
which again connects data between both pictures. Knowing that spherical
harmonics can be expressed in terms of associated Legendre functions
$P^m_j$ by
\be Y^m_j(\theta,\phi)=e^{im\phi}P^m_j(\cos\theta),\ee whose
value at the equator $\theta_0=\pi/2$ is given in terms of Gamma functions
as
\be P^m_j(0)=\frac{2^m}{\sqrt{\pi}}\cos\left(\um
\pi(j+m)\right)\frac{\Gamma(\um j+\um m+\um)}{\Gamma(\um j-\um m+1)},\ee
we immediately realize that ${\cal Y}(\pi/2)_{kn}=0$ for $n$ odd. In other
words, for $\theta_0=\pi/2$, the reconstruction process in the Euler angle
picture fails unless we restrict to the subspace of functions $\psi$ with
null odd Fourier coefficients (i.e.,  $a_{n-j}=0$ for $n$ odd).

\appendix

\section{Rectangular Fourier Matrices}
\label{RFM}

Let $N,M\in \mathbb{N}$, and let ${\cal F}_{NM}$ be the $N\times M$ matrix
\be
({\cal F}_{NM})_{nm}=\frac{1}{\sqrt{N}} e^{-i2\pi nm/N}\,,\,n=0,1,\ldots,N-1\,,\,m=0,1,\ldots,M-1\,.
\ee

We shall denote these matrices Rectangular Fourier Matrices (RFM). For $N=M$ we recover the
standard Fourier matrix ${\cal F}_{N}$. Let us study the properties of these matrices in the
other two cases, $N>M$ and $N<M$.

\subsection{Case $N>M$}

The case $N>M$ is the one corresponding to oversampling, and it is the easiest one,
since it is very similar to the $N=M$ case. Let us first introduce some notation.

Let $\iota_{NM}: \mathbb{C}^M\rightarrow \mathbb{C}^N$ be the inclusion into the
first M rows of $\mathbb{C}^N$ (i.e., padding a $M$-vector with zeros). And let
$p_{MN}: \mathbb{C}^N\rightarrow \mathbb{C}^M$ the projection onto the
first $M$ rows of $\mathbb{C}^N$ (i.e., truncating a $N$-vector). The matrix expression for this
applications are
\be
\iota_{NM}=\left( \ba{cccc} 1 & 0 & \ldots & 0\\ 0 & 1 & \ldots & 0 \\ \vdots & \vdots & \ddots &
\vdots \\ 0 & 0 & \ldots & 1 \\ \vdots & \vdots & \vdots & \vdots \\ 0 & 0& \ldots & 0 \ea \right)\,,
\qquad
p_{MN} = \left( \ba{cccccc} 1 & 0 & \ldots & 0 &\ldots & 0\\ 0 & 1 & \ldots & 0 &\ldots & 0\\
 \vdots & \vdots & \ddots &\vdots & \vdots & \vdots \\
 0 & 0 & \ldots & 1 & \ldots & 0 \ea \right).
\ee

It can be easily checked that
\be
p_{MN} = (\iota_{NM})^* \,,\qquad  p_{MN}{\,} \iota_{NM}=I_M\,,\qquad \iota_{NM}{\,} p_{MN} = P_M\equiv
\left( \ba{c|c} I_M & 0 \\ \hline  0 & 0\ea \right)_{N\times N}
\ee
where $I_M$ stands for the identity matrix in $\mathbb{C}^M$.

Given the square matrices $A$ and $B$ acting on $\mathbb{C}^N$ and $\mathbb{C}^M$, respectively,
we define the square matrices $A^\downarrow$ and $B^\uparrow$ through the commutative diagrams
\be
\begin{CD}
\mathbb{C}^M @>A^{\downarrow}>> \mathbb{C}^M \\
@V\iota_{NM}VV @AAp_{MN}A \\
\mathbb{C}^N @>A>> \mathbb{C}^N
\end{CD} \qquad\qquad\qquad\qquad
\begin{CD}
\mathbb{C}^N @>B^\uparrow>> \mathbb{C}^N \\
@Vp_{MN}VV @AA\iota_{NM}A \\
\mathbb{C}^M @>B>> \mathbb{C}^M
\end{CD}
\ee

The  matrix $A^{\downarrow} =p_{MN} {\,} A {\,} \iota_{NM} $ is the truncation of $A$ to a $M\times M$ matrix and the
matrix $B^\uparrow= \iota_{NM} {\,} B {\,} p_{MN} \equiv
\left( \ba{c|c} B & 0 \\ \hline  0 & 0\ea \right)_{N\times N}$ is the padded version of $B$. Also
note
that $P_M=(I_M)^\uparrow$.

{}From these definitions the following properties of the Rectangular Fourier Matrices are
derived:
\bea
{\cal F}_{NM}&=&{\cal F}_N{\,} \iota_{NM}\,,\qquad {\cal F}_{NM}^* = p_{MN}{\,}{\cal F}_N^*
\nn\\
{\cal F}_{NM}^*{\,} {\cal F}_{NM} &=& I_M \,,\qquad  {\cal F}_{NM}{\,} {\cal F}_{NM}^* =
{\cal F}_N {\,} P_M {\,} {\cal F}_N^*\,.
\eea

\subsection{Case $N<M$}

The case $N<M$ is the one corresponding to undersamplig, and it is not as easy as the $N>M$ case.
The same definitions as in the previous case also apply here, but interchanging the roles of
$N$ and $M$. Thus, in this case we have
\be
p_{NM}{\,} \iota_{MN}=I_N\,,\qquad \iota_{MN}{\,} p_{NM} = P_N\equiv
\left( \ba{c|c} I_N & 0 \\ \hline  0 & 0\ea \right)_{M\times M} = (I_N)^\uparrow
\ee

The RFM in this case now read
\bea
{\cal F}_{NM} &=& \left( \ba{c|c|c|c|c} {\cal F}_N &{\cal F}_N & \stackrel{q\ {\rm times}}{\ldots} &
{\cal F}_N & {\cal F}_{Np} \ea \right)
\nn \\
{\cal F}_{NM}^* &=& \left( \ba{c} {\cal F}_N^* \\ \hline {\cal F}_N^* \\ \hline\vdots\,
\hbox{\footnotesize{ q\ {\rm times}}}
\\ \hline {\cal F}_N^* \\ \hline {\cal F}_{Np}^* \ea \right)
\eea
where $p = M\  {\rm mod}\  N$ and $q= M\  {\rm div}\  N$.

Instead of working with these matrices, it is more convenient to ``complete" them so
as to have an integer multiple of Fourier matrices. Let $\bar{M}$ be the smaller multiple of $N$
greater or equal to $M$, and $\bar{q}=\bar{M}/N$ the ceiling of $M/N$. Note that
\be
\bar{q}=\left\{ \ba{cc} q & p=0 \\ q+1 & p\neq 0 \ea \right.
\ee

Then
\be
{\cal F}_{N\bar{M}} =\left( \ba{c|c|c|c} {\cal F}_N &{\cal F}_N & \stackrel{\bar{q}\ {\rm times}}{\ldots} &
{\cal F}_N \ea \right)\,,
\ee
a similar expression for ${\cal F}_{N\bar{M}}^*$ is obtained, and
\bea
{\cal F}_{NM} &=& {\cal F}_{N\bar{M}} {\,} \iota_{\bar{M}M} \,,\qquad
{\cal F}_{NM}^* = p_{M\bar{M}} {\,} {\cal F}_{N\bar{M}}^* \nn \\
{\cal F}_{NM} {\,} {\cal F}_{NM}^* &=& q I_N + {\cal F}_N {\,} P_p {\,} {\cal F}_N^*\,,
\qquad
 {\cal F}_{NM}^* {\,} {\cal F}_{NM} = (\hat{I}_{\bar{M}})^\downarrow \equiv \hat{I}_M
\eea
where
\be
\hat{I}_{\bar{M}} = \left( \ba{c|c|c|c} I_N & I_N & \stackrel{\bar{q}\ {\rm times}}{\ldots} & I_N \\
\hline I_N & I_N & \ldots & I_N \\
\hline \vdots\,\hbox{\footnotesize{ $\bar{q}$\ {\rm times}}} & \vdots & \ddots & \vdots \\
\hline I_N & I_N & \ldots & I_N \ea \right)_{\bar{M}\times\bar{M}}
\ee

\section{Circulant Matrices}
\label{Circulantes}

The overlapping kernel operator $\cb$ has a circulant matrix structure which
gives a deep insight into the process taking place and we may take advantage of this fact
to diagonalize it in the case of undersampling where RFM are more difficult to
handle.

Indeed, note that
\be {\cal B}_{kl}=\langle
z_k|z_l\rangle=\frac{1}{2^{2s}}\left(1+e^{2\pi
i(l-k)/N}\right)^{2s}=\cc_{l-k},\; k,l=0,\ldots,N-1.
\label{resolentriesunder}\ee
where $\cc_n = \frac{1}{2^{2s}}\left(1+e^{2\pi
in/N}\right)^{2s}$, which shows a circulant matrix structure
\be {\cal B} ={\rm circ}(\cc_0,\cc_1,\ldots,\cc_{N-1})=\left(
\begin{array}{cccc}
  \cc_0 & \cc_1 & \ldots & \cc_{N-1} \\
  \cc_{N-1} & \cc_0 & \ldots & \cc_{N-2} \\
  \vdots & \vdots & \ddots & \vdots \\
  \cc_1 & \cc_2 & \ldots & \cc_0 \\
\end{array}
\right)= \sum_{j=0}^{N-1} \cc_j \Pi^j\equiv P_c(\Pi),\ee
where
$$\Pi=\left(%
\begin{array}{cccc}
  0 & 1 & \ldots & 0 \\
  \vdots & \vdots & \ddots & \vdots \\
  \vdots & \vdots & \ldots & 1 \\
  1 & 0 & \ldots & 0 \\
\end{array}%
\right),\qquad (\Pi^N=I_N,\quad\Pi^t=\Pi^*=\Pi^{-1}=\Pi^{N-1}), $$ is the
\emph{generating matrix} of the circulant matrices and $P_c(t)$ is the
\emph{representative polynomial} of the circulant (we put $\Pi^0\equiv
I_N$). According to the general theory (see e.g.\cite{circulante}), every
circulant matrix is diagonalizable, whose eigenvectors are the columns of
the Vandermonde matrix $V_N=V(z_0,\ldots,z_{N-1})= \sqrt{N}\cf^*$ and
whose eigenvalues $\hat\lambda_k$ can be computed through its
representative polynomial as\footnote{We are using the complex conjugate
$\bar{z}_k$ instead of $z_k$ in the representative polynomial to obtain
the more convenient factorization ${\cal B} ={\cal F}_N D{\cal F}_N^*$
instead of the standard one ${\cal B} ={\cal F}_N^* D'{\cal F}_N$, where
the eigenvalues in $D'$ have reversed order with respect to those of
$D$.}
\bea \hat\lambda_k&=&P_c(\bar{z}_k)=\sum_{l=0}^{N-1}\cc_lz_k^{-l}
=2^{-(2s)}\sum_{l=0}^{N-1}\sum_{n=0}^{2s} \tbinom{2s}{n} e^{2\pi i
l(2s-n)/N} e^{-2\pi i k l/N}\nn\\
&=&\frac{N}{2^{2s}}\sum_{l=0}^{\bar{q}-1}\tbinom{2s}{k+lN},\;k=0,\ldots,N-1,
\label{eigenvaluesunder}\eea
where $\bar{q}$ is the ceiling of $(2s+1)/N$ and we have used the
orthogonality relation (\ref{exponencial}), although in this case, since
$N\leq 2s+1$ there can be more terms in the sum. All of them are strictly
positive, and it is easy to proof that ${\cal B} ={\cal F}_N \hat{D} {\cal
F}_N^*$, where $\hat{D}={\rm
diag}(\hat{\lambda}_0,\ldots,\hat{\lambda}_{N-1})$. $\blacksquare$

\section*{Acknowledgements}

Work partially supported by the MCYT and Fundación Séneca under projects
FIS2005-05736-C03-01 and 03100/PI/05

\footnotesize
\centerline{\rule{9pc}{.01in}}
\bigskip
\centerline{Departamento de Matemática Aplicada y Estad\'\i stica,
Universidad Politécnica de Cartagena}
\centerline{Paseo Alfonso XIII 56, 30203
Cartagena, Spain}
\centerline{e-mail:Manuel.Calixto@upct.es}
\medskip
\centerline{Departamento de Matemática Aplicada,
Universidad de Murcia, Facultad de Informática}
\centerline{Campus de Espinardo, 30100
Murcia, Spain }

\centerline{e-mail: juguerre@um.es}
\medskip
\centerline{Departamento de Matemática Aplicada y Estad\'\i stica,
Universidad Politécnica de Cartagena}
\centerline{Paseo Alfonso XIII 56, 30203
Cartagena, Spain}
\centerline{e-mail: JCarlos.Sanchez@upct.es}
\medskip

\end{document}